\newcommand{\un}[1]{\ensuremath{\, \mathrm{#1}}}

\documentclass[preprint]{aastex}
\usepackage{natbib}
\usepackage{graphicx, emulateapj5, apjfonts}
\usepackage{onecolfloat}
\usepackage{multirow}
\usepackage{epsf}
\usepackage{cite}
\usepackage{subfigure}
\RequirePackage{lineno}

\shorttitle{Searches for High-Energy Neutrino Emission in the Galaxy} 
\shortauthors{R.~Abbasi et al.} 

\journal{}
\begin{document}

\title{Searches for High-Energy Neutrino Emission in the Galaxy with the Combined IceCube-AMANDA Detector}
\author{
 IceCube Collaboration:
 R.~Abbasi\altaffilmark{1},
 Y.~Abdou\altaffilmark{2},
 M.~Ackermann\altaffilmark{3},
 J.~Adams\altaffilmark{4},
 J.~A.~Aguilar\altaffilmark{5},
 M.~Ahlers\altaffilmark{1},
 D.~Altmann\altaffilmark{6},
 K.~Andeen\altaffilmark{1},
 J.~Auffenberg\altaffilmark{1},
 X.~Bai\altaffilmark{7,8}
% ...
 }
 \altaffiltext{1}{Dept.~of Physics and Wisconsin IceCube Particle Astrophysics Center, University of Wisconsin, Madison, WI 53706, USA}
 \altaffiltext{2}{Dept.~of Physics and Astronomy, University of Gent, B-9000 Gent, Belgium}
 \altaffiltext{3}{DESY, D-15735 Zeuthen, Germany}
 \altaffiltext{4}{Dept.~of Physics and Astronomy, University of Canterbury, Private Bag 4800, Christchurch, New Zealand}
 \altaffiltext{5}{D\'epartement de physique nucl\'eaire et corpusculaire, Universit\'e de Gen\`eve, CH-1211 Gen\`eve, Switzerland}
 \altaffiltext{6}{Institut f\"ur Physik, Humboldt-Universit\"at zu Berlin, D-12489 Berlin, Germany}
 \altaffiltext{7}{Bartol Research Institute and Department of Physics and Astronomy, University of Delaware, Newark, DE 19716, USA}
 \altaffiltext{8}{Physics Department, South Dakota School of Mines and Technology, Rapid City, SD 57701, USA}
% ...

\begin{abstract}

We report on searches for neutrino sources at energies above 200~GeV in the Northern sky of the galactic plane, using the data 
collected by the South Pole neutrino telescope IceCube and AMANDA.
The galactic region considered in this work includes the Local Arm towards the Cygnus region and our closest approach to the Perseus Arm.
The searches are based on the data collected between 2007 and 2009. During this 
time AMANDA was an integrated part of IceCube, which was still under construction and operated 
with 22-strings (2007-8) and 40-strings (2008-9) of optical modules deployed in the ice. 
By combining the advantages of the larger IceCube detector with the lower energy threshold of the more compact AMANDA detector, we obtain an improved sensitivity at energies below $\sim$10~TeV with
respect to previous searches. The analyses presented here are: a scan for point sources within the galactic plane; a search optimized for multiple and extended sources in the Cygnus region,
which might be below the sensitivity of the point source scan; and studies of seven pre-selected neutrino source candidates. For one of them, Cygnus X-3, a time-dependent search for neutrino emission
in coincidence with observed radio and X-ray flares has been performed. No evidence of a signal is found, and upper limits are reported for each of the searches. 
We investigate neutrino spectra proportional to E$^{-2}$ and E$^{-3}$ in order to cover the entire range of possible neutrino spectra.
The steeply falling E$^{-3}$ neutrino spectrum can also be used to approximate neutrino energy spectra with energy cutoffs below 50~TeV since these result in a similar energy distribution of events in the detector.
For the region of the galactic plane visible in the Northern sky, 
the 90\% confidence level muon neutrino flux upper limits are in the range E$^3$dN/dE$\sim 5.4 - 19.5 \times 10^{-11}~\rm{TeV^{2} cm^{-2} s^{-1}}$ for point-like neutrino sources in the energy region [180.0~GeV - 20.5~TeV]. 
These represent the most stringent upper limits for soft-spectra neutrino sources within the Galaxy reported to date. 
\end{abstract}

\keywords{acceleration of particles, cosmic-rays, neutrinos}

\section{Introduction}

The IceCube neutrino telescope at the South Pole has successfully been completed in December 2010. 
IceCube is the most sensitive telescope to date to search for high-energy neutrino sources, whose existence is intimately related to the acceleration of hadrons and their interaction in the environment of their accelerator. The interaction of high-energy protons and nuclei with ambient matter or radiation leads to the generation of both gamma-rays and neutrinos of similar energy \citep{2006PhRvD..74c4018K, 2008PhRvD..78c4013K}. However, it is difficult to infer the contribution of a possible hadronic component from the observed gamma-rays, since gamma-ray emission can also be produced by relativistic electrons via Inverse Compton scattering. 
Moreover, the most energetic gamma-rays have a high probability to be absorbed on their way to Earth, and the observed spectra may,
after successive absorption and emission processes \citep{Moskalenko06}, not be the same as the primary spectra.
The detection of a flux of high-energy neutrinos from astrophysical sources, even if challenging, can thus provide unique insights into the acceleration mechanisms and the origin of cosmic-rays.

The IceCube neutrino telescope has a full-sky field of view at any time and has thus the potential to observe neutrino point sources at any position in the sky albeit with different discovery potential depending on the source location and the neutrino energy spectrum of the source. For a source following an E$^{-2}$ spectrum in the energy range from 1~TeV up to a few PeV IceCube can discover high-energy neutrino sources at the flux level of $10^{-11}-10^{-12}$~$\rm{erg~cm^{-2}~s^{-1}}$ \citep{2011ApJ...732...18A} if the source location is known from other observations. Assuming isotropic emission, this translates to source luminosities
of $L (\rm{E > 1~TeV}) \sim 10^{33}~\rm{erg~s^{-1}}$ for a source at a distance of 2~kpc. A search for neutrino point sources at any location in the sky with the IceCube 40-strings detector has been presented in \citep{2011ApJ...732...18A} using E$^{-2}$ and flatter spectra for the optimization of the analysis. This work focusses on the more specific case of Galactic neutrino sources and the energy spectra associated to them.

Among the most promising candidate sources of cosmic-rays in the Galaxy are the remnants of supernovae (both shell-type and pulsar wind nebulae), the jets of microquasars, and the collective winds of massive
stars \citep{hillas05, AGILE_cygx3, fermi_cygx3, HESS_Wd2, HESS_Wd1, HESSUidMassiveClus}. Due to the large amount of energy released in a supernova explosion ($\sim 10^{51}$~erg),
supernova remnants are prime candidates to be sources of Galactic cosmic-rays. In microquasars, the kinetic energy carried by the jet accounts for at least $10^{36}~\rm{erg~s^{-1}}$,
inferred from the observed non-thermal luminosities \citep{Gallo05_cygx1, Margon84_JetPower}. The total energy injected into the interstellar medium by the winds of OB and
Wolf-Rayet stars can be as high as $\sim 10^{39}~\rm{erg~s^{-1}}$, like in the case of the Cygnus OB2 association \citep{CygOB2Power}; and levels of $\sim10^{38}~\rm{erg~s^{-1}}$
can be achieved by a single young pulsar. What remains undetermined is the fraction of total energy per source that goes into cosmic-ray acceleration, as well as the probability
for the interaction of cosmic-rays close to their source. The observation of the products of cosmic-ray interactions, i.e. gamma-rays and neutrinos, can shed light on these still unsolved problems.

The highest energies (E > 100~TeV) are only accessible by means of extensive air-shower (EAS) arrays, in the case of gamma-rays, and $\rm{km^3}$-volume neutrino detectors like IceCube.
Results from Milagro \citep{MgroCygnus} and ARGO-YBJ \citep{ARGOCygnus}, demonstrate that 
the gamma-ray emission is faint at very high energies \citep{CasaMia98}. However, these gamma-ray observations do not impose constraints on neutrino production,
due to internal and external absorption of gamma-rays at the energies considered.

Most of our knowledge of gamma-ray sources comes from Cherenkov telescopes like H.E.S.S., MAGIC, and VERITAS, working in the energy range 100~GeV - 100~TeV.
In the past few years, a large family of Galactic accelerators have been observed to have the bulk of their gamma-ray emission at
energies below 50~TeV \citep{2006A&A...448L..43A, 2009A&A...503..817A, 2007ApJ...664L..87A, 2007A&A...474..937A} and/or to be softer than the $\rm{dN/dE \propto E^{-2}}$ spectrum that is generally
expected from first-order Fermi shock acceleration \citep{1949PhRv...75.1169F, 1954ApJ...119....1F}. 
Many of these sources reside relatively nearby, and external gamma-ray absorption in the interstellar radiation field is not likely.
If the detected gamma-rays are related to hadronic acceleration and are produced in transparent sources, the expected distribution of neutrino energies
has the same spectral index and a lower energy cutoff than the gamma-ray spectrum \citep{2006PhRvD..74c4018K}.
The modeling of cosmic-ray sources with diffusive shock acceleration also allows for the presence of spectra steeper or flatter than the generic E$^{-2}$ behavior,
depending on the configuration of the shock \citep{Bell78, Bell78b, Schlickeiser89, Schlickeiser89b, Meli08}.
The influence of diffusion in the sources themselves may modify the spectra to produce primary spectra of E$^{-2.3}$ or steeper \citep{Biermann09, Biermann2010}. 
In order to target soft-spectra sources, we have optimized the search here reported for a generic spectrum proportional to E$^{-3}$.

\begin{figure}[t]
\begin{center}
\includegraphics[width=0.4\textwidth]{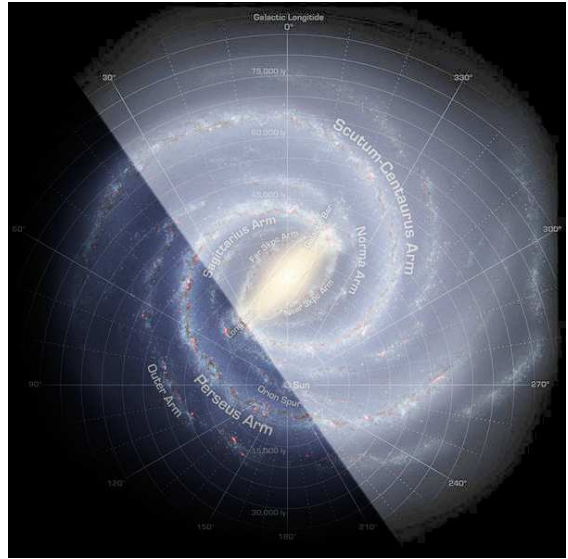}
\caption{Artistic rendering of the Milky Way made from optical, IR and radio data from \citep{2009PASP..121..213C}.
The part of the galaxy within the field of view of the IceCube analyses in this paper is from galactic longitude $40^\circ < \ell < 210^\circ$ (i.e. lower left region).}
\label{Galaxy}
\end{center}
\end{figure}

In this work, we use the 22- and 40-strings configurations of IceCube (IC22, IC40) as well as the Antarctic Muon And Neutrino Detector Array (AMANDA) (IC22+A, IC40+A) to enhance the sensitivity for soft-spectra sources,
or sources with energy cut-offs. We have used AMANDA as an integrated low-energy extension of IC22 and IC40 and developed an analysis strategy that is optimized for a high retention of signal events below 10~TeV.
We have used the resulting low-energy optimized data samples to search for Galactic neutrino emission above $\approx$200~GeV. At these energies, IceCube's field of view covers
the range of Galactic longitude $40^\circ < \ell < 210^\circ$, as illustrated in Figure~\ref{Galaxy}. The performed searches include a scan of the accessible part of the Galactic Plane, a dedicated analysis of
the Cygnus region, the search for neutrino emission from a pre-defined list of interesting astrophysical objects and an analysis that searches for time-dependent neutrino emission from Cygnus X-3 in correlation
with radio flares. The paper is organized as follows: Section 2 describes the relevant technical aspects of IceCube, of AMANDA and of its integration into IceCube. 
Section 3 reports about the analysis methods that have been applied and the respective astrophysical targets. Section 4 explains the details and characteristics of the obtained data samples,
and Section 5 provides the results. 

\section{The Combined Detector: IceCube and AMANDA}

\subsection{IceCube}

During the construction phase from 2004 to 2010, the operational
configuration of IceCube increased year by year (see Figure~\ref{IC_AM})
to finally cover a volume of approximately one cubic kilometer.
IceCube, including its DeepCore extension, is composed of 86 strings each holding 60 digital optical modules (DOMs).
Each DOM contains a 10-inch photomultiplier tube (PMT) and an on-board 
signal read-out and digitization system, all housed in a glass pressure vessel \citep{DAQpaper}.
78 of the 86 strings in the array form a hexagonal grid
with a typical distance of 125~m between neighboring strings. The vertical distance between DOMs on the same string is 17~m.
The remaining 8 strings are part of the low-energy extension DeepCore \citep{DCPaper} and are deployed
in the deepest, clearest ice at the center of the detector with a smaller vertical and horizontal spacing between the DOMs of 7~m and 60~m, respectively. 
The DOMs detect Cherenkov radiation emitted by 
secondary charged particles produced in interactions of high-energy neutrinos with nuclei in the
ice or the bedrock below the ice. To enhance the detection of
light from upward-going particles, the PMTs point downwards. 
In order to
avoid a deterioration of the analog PMT signal, the signal is digitized directly in the DOMs
with a set of Analog Transient Waveform Digitizers (ATWDs) and a Fast Analog to Digital
Converter (fADC) \citep{2010NIMPA.618..139A}. 
The events that are used in this analysis are selected by a multiplicity trigger which requires at least 8 hit DOMs within a time window of $5~\mu$sec.
The DOMs send their recorded signals to the surface and an event is constructed if the trigger conditions are met. 
In the detector configurations used in this work, only
DOMs for which there is also a signal from one of the nearest two DOMs
above or the nearest two DOMs below within
1 $\mu$sec (so-called Hard Local Coincidence) are considered in the
trigger and the event building to suppress noise contributions. 
An event contains all DOM readouts associated with the trigger as well as all further readouts within $\pm 10~\mu$sec around the trigger time.

\subsection{AMANDA}

\begin{figure}[t]
\begin{center}
\includegraphics[width=0.7\textwidth]{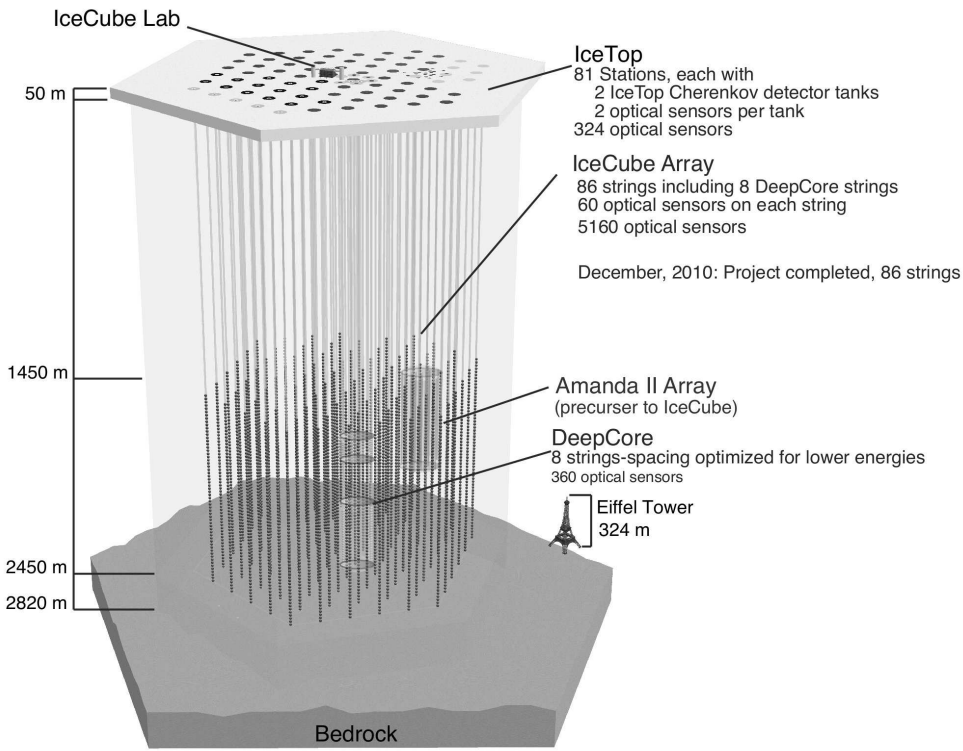}
\caption{View of the IceCube array. 
AMANDA is completely surrounded by IceCube strings and presents a more compact structure. }
\label{IC_AM}
\end{center}
\end{figure}

After a construction phase from 1993 until 2000, the completed AMANDA-II
detector took data as a standalone neutrino telescope from February 2000 until December 2006. This
configuration consisted of 677 optical modules (OMs) on 19 strings.
Most of the optical modules were deployed at depths between 1500 and 2000~m whereas
IceCube extends down to 2450~m.
For data analysis, a total of 526 OMs have been used. The AMANDA strings follow a 
roughly cylindrical geometry, as is shown in Figure~\ref{IC_AM}.

The typical distance between adjacent strings is around 40~m and the average vertical spacing between the modules is about 15m.
From February 2007 until April 2009, AMANDA was operated as an integrated
part of IceCube.
In many aspects, IceCube is technologically more advanced than AMANDA,
reflecting general progress as well as experience collected during the
operation of AMANDA \citep{2006NIMPA.556..169C}.
In particular, the signal transfer from the optical modules to the surface is
different. As mentioned above, IceCube DOMs digitize the PMT signal directly in the
ice. They also generate the HV for the PMTs
in the DOMs. In contrast, 
AMANDA OMs produced analog signals that were sent to the AMANDA data acquisition
system which was located in the Martin A. Pomerantz Observatory (MAPO).
The original data acquisition system (DAQ) could register the leading and trailing edge time of up
to 8 pulses per OM per event and 
only the total charge.
The same cables were used to transfer the analog data and to provide the HV to the PMTs. 
For AMANDA strings 11 to 19, an additional
connection via optical fibers was installed to transmit the PMT signals with a better time
resolution. Moreover, AMANDA string 18 \citep{2006NIMPA.556..169C}
was equipped with prototypes for the IceCube DOMs,
including the capability of on-board waveform digitization. This option
however was used only in testing mode and was not included in the data acquisition schemes
used for physics analysis.

The AMANDA DAQ was upgraded starting from 2002.
Flash ADC modules called Transient Waveform Recorders (TWRs) were
installed in the new DAQ in order to digitize the
analog waveforms from the AMANDA OMs at the surface.
The upgraded data acquisition operated in parallel to the analog one until 2006.
From 2007 on, only the TWR-DAQ was operational.
This surface waveform digitization stored more information
such as the induced charge and the arrival times of individual pulses. As the TWR-DAQ was also faster
than the previous DAQ, the trigger threshold could be reduced. Trigger thresholds from 8 to 13
hit OMs were used
during different years. 
This upgrade significantly improved the performance of AMANDA and ultimately allowed to use it to enhance IceCube's performance at low energies
\citep{ICWIMPs}.

\subsection{AMANDA as an integrated part of IceCube}

Since AMANDA is about eight times more densely instrumented than IceCube and fully surrounded by IceCube strings, it offered the potential to increase the low-energy performance of IceCube
and to be used as the first low-energy core inside a large neutrino telescope. This lead to its integration in the data taking of IceCube. 

The operational integration of AMANDA into IceCube required the establishment of connections between the two detectors
for the exchange of trigger information to be able to merge events as well as for an accurate synchronization in time.
MAPO is about 300~m away from the IceCube Control Lab (ICL)
that houses the IceCube surface data acquisition. 
Optical fibers have been used in order to connect the two buildings. 
Moreover, a TCP/IP connection was
established for the communication between the buildings. 
A GPS module was installed to synchronize the TWRs
responsible for the digitization of the AMANDA waveforms
and to synchronize the detector with IceCube. 
The IceCube clock was used as reference.
An optical fiber connection was used to
transmit the AMANDA trigger
signal to IceCube. 
In the integrated mode, AMANDA and IceCube were still triggered separately.
Since AMANDA had a lower energy threshold than IceCube, 
a readout of IceCube was initiated every time AMANDA triggered, 
even if there were not sufficient hits in IceCube to produce a trigger by itself. 
AMANDA was not read-out in correspondence with IceCube triggers.
Events from AMANDA and IceCube were merged on the basis of a time coincidence. 
As the duration of AMANDA records was fixed to 10.24~$\mu$s, 
while the duration of IceCube records was extended if new triggers 
occurred within the readout window there was a possibility that more than one record 
in AMANDA was associated with one record in IceCube.
In this case, they were all included in the same combined event.

\section{Methods and Targets}

The IceCube neutrino telescope monitors the entire sky without the need of explicit pointing. The energy- and zenith-dependent sensitivity of the IceCube 40-strings configuration are described 
in \citep{2011ApJ...732...18A}. In previous works, a generic, unbroken $\rm{dN/dE \propto E^{-2}}$ signal spectrum up to the PeV 
region has been assumed for the optimization of the data analysis and the evaluation of the detector performance \citep{2011ApJ...732...18A}. 
This approach achieves the best signal to noise in the energy range above a few TeV, since the assumed signal spectrum is significantly 
harder than the characteristic spectrum of (background) atmospheric neutrinos $\rm{dN/dE \propto E^{-3.7}}$.
However, a lower energy threshold is of primary importance for the search for neutrino sources characterized by an energy cutoff or by soft spectra.
In the optimization of the analysis, we have used a generic, soft power-law spectrum following $\rm{dN/dE \propto E^{-3}}$ 
and we have also considered the Crab nebula spectrum measured by H.E.S.S. \citep{HESS_crab} which would correspond to a neutrino spectrum $\rm{dN/dE \propto E^{-2.4}}$ with an exponential cutoff at 7~TeV
(provided that all the measured gamma-rays are of hadronic nature). 
This last spectrum is representative of a "low-energy" source and its study is very instructive in the understanding of the impact of an energy cutoff on the performance of IceCube.
We will refer to this spectrum throughout the paper as a "Crab-like" spectrum. 
We report in the following on the different searches that have been performed.

\subsection{Galactic Plane Scan and Source List}

The location of the Solar System in the local spiral arm gives us a particular view of the Galaxy. Given the
vertical scale of the thin disk and the distribution of cold gas in the Galaxy, we see most of the galactic accelerators
projected in a narrow band close to the Galactic Plane. In the energy range of interest for the detection of galactic neutrino sources, IceCube can explore the Northern sky,
which includes part of the first quadrant of the Galaxy, the whole second quadrant, and a small portion of the third Galactic quadrant.
The search for neutrino sources in the Galactic Plane is performed by superimposing a fine grid over the region of the sky within the Galactic coordinates $36^{\circ} < \ell < 210^{\circ}$, $-5^{\circ} < b < 5^{\circ}$.
The step size of the grid is chosen to be smaller than the angular resolution achieved in the analysis reported in Section 4.4.2. 

In the analysis of IC22+A, a grid of $0.5^{\circ} \times 0.5^{\circ}$ has been used, while a $0.25^{\circ} \times 0.25^{\circ}$ grid has been chosen for the IC40+A, given the improved angular resolution
(see Figure~\ref{ANG_RES}). At each point of the grid, an unbinned maximum likelihood ratio test is performed on all selected events. The likelihood of a composite signal and background hypothesis is compared
to the background-only hypothesis, similar to the method described in \citep{2008APh....29..299B}, without the inclusion of an energy term in the likelihood. 

Seven particularly interesting sources have been studied individually in this analysis as representatives of different types of Galactic accelerators. The interest in these sources is motivated by the observation
of a GeV-TeV gamma-ray counterpart at the time of the analysis. For Cygnus X-3, due to the high variability in the radio and X-ray bands, we have tested the hypothesis of variable neutrino emission and performed
a time-dependent analysis. The other sources are listed below and are treated as steady point source candidates. 

{\bf Crab nebula, distance:$\approx$2.0~kpc} \citep{1968AJ.....73..535T}: The Crab nebula is powered by a pulsar with a spin-down luminosity of $\sim 5\times 10^{38}~\rm{erg~s^{-1}}$. This energy is injected into
relativistic particles and magnetic fields \citep{1984ApJ...283..710K}, although the exact composition of the pulsar wind, as well as the mechanism by which the total power of the pulsar is transported and dissipated
is not known. It is an efficient particle accelerator, where $\approx 60\%$ of the total power of the pulsar is injected into relativistic electrons which emit synchrotron radiation from radio to X-rays \citep{Hester08}.
Although it appears as the strongest gamma-ray source in the sky, the ratio between the gamma-ray luminosity at E $>$ 1~TeV and the spin-down luminosity is of the order of $10^{-5}$ \citep{HEGRA_crab, HESS_crab}.
The simplest interpretation of this is that electrons rapidly lose their energy through synchrotron radiation at lower frequencies, and that the majority of cosmic-rays, if present in a significant proportion,
escape from the source without interaction. The constraint on the steady neutrino production in the Crab obtained by IceCube \citep{FlarePap} is at the level of $L_{\nu} \lesssim 2 \times 10^{35}~\rm{erg~s^{-1}}$,
a factor of $\approx 3.4$ larger than the luminosity in gamma-rays assuming the H.E.S.S. gamma-ray spectrum \citep{HESS_crab} (and its corresponding neutrino spectrum expected in case of an hadronic origin)
extrapolated to the energy range between 400~GeV and 40~TeV. 
 
{\bf Cas A, distance:$\approx$3.4~kpc } \citep{1995ApJ...440..706R}: This source is a classical shell-type Supernova Remnant (SNR). Its high-energy gamma-ray flux was detected by HEGRA in the energy region between
1~TeV to 10~TeV without any evidence of an energy cutoff \citep{HEGRA_casA}, and detected by MAGIC down to 250~GeV following a power-law spectrum $\propto E^{-2.3}$ and with an integrated photon flux above 1 TeV of
$\approx 7.3 \times 10^{-13}~\rm{cm^{-2}~s^{-1}}$ \citep{MAGIC_CasA}.

{\bf IC~443, distance:$\approx$1.5~kpc} \citep{1984ApJ...281..658F}: IC~443 is an asymmetric shell-type SNR, where part of the shell is impacting on a molecular cloud, accelerating particles to very high energies
in the process. TeV gamma-rays are observed arriving from the molecular line emission region, giving support to a hadronic origin of the TeV gamma-rays.
The spectrum measured in the energy range from 100~GeV to 1.6~TeV is well fitted by a very steep power-law $\propto \rm{E^{-3.1}}$ \citep{MAGIC_ic443}.
The integrated photon flux above 1 TeV obtained by extrapolation is $\approx 3.2 \times 10^{-13}~\rm{cm^{-2}~s^{-1}}$.

{\bf W51, distance:$\approx$6.0~kpc} \citep{1967AnAp...30...59K}: This source has been detected in GeV gamma-rays by the Fermi-LAT telescope \citep{FermiW51C}, at TeV energies by H.E.S.S. \citep{HESSW51}
and by MAGIC \citep{MAGICW51}. The high-energy emission is thought to arise from the interaction between a composite SNR (W51C) with a molecular cloud present in the region.
The high luminosities observed in GeV gamma-rays, greater than $10^{36}~\rm{erg~s^{-1}}$, and the hints of a hadronic origin for the gamma-ray spectrum makes this an interesting target for IceCube despite
its large distance. The MAGIC Collaboration recently extended the spectrum from the highest Fermi/LAT energies to 5~TeV and finds that the spectral index of the source follows a single
power law with an index of $2.58 \pm 0.07_{stat} \pm 0.22_{syst}$ \citep{MAGICW51}.

{\bf LS~I+61~303, distance:$\approx$2.0~kpc} \citep{1991AJ....101.2126F}: This source is a high-mass X-ray binary with a compact object in an eccentric orbit around a Be star. 
The nature of the compact object is not known, and both a pulsar wind model and a microquasar model
have been suggested for this source. MAGIC detected very high energy emission modulated with the orbital period \citep{magicLSI}. 
The highest significant detection is obtained around apastron, at orbital phases 0.6-0.7, 
with a spectrum following dN/dE $\approx 2.6\times 10^{-12} E^{-2.6} ~\rm{TeV^{-1}~cm^{-2}~s^{-1}}$, at E $>$ 300~GeV.
No TeV emission is observed at periastron, although significant gamma-ray absorption in the strong radiation
field of the Be star is expected in this case \citep{Sierpowska09}. 
This scenario is supported by the detection of MeV-GeV gamma-rays by the Fermi 
satellite \citep{FermiLSI}, which may result from the cascade process in $\gamma\gamma \rightarrow e^{+}e^{-}$. 
The hypothesis of particle injection along the whole orbit is then a plausible option.
This, together with the
considerable amount of both matter and radiation
from the companion star available for cosmic ray interactions, makes this
source an interesting candidate for steady neutrino emission. 
For the search of periodic neutrino emission from binary systems
performed by IceCube, we refer to \citep{MikePaper}.

{\bf SS~433, distance:$\approx$5.0~kpc} \citep{1987ApJ...321..822R}: SS~433 is a confirmed microquasar and a black hole candidate in orbit around a massive star. 
The source exhibits two oppositely directed relativistic jets which are thought to eject material at a rate larger 
than $10^{-6}~\rm{M_{\odot}~year^{-1}}$ \citep{Begelman80}. 
It is the only X-ray binary system in which hadrons have been found in the jet \citep{2002Sci...297.1673M}. 
The entire source is embedded within a nebulous structure (W50) which is thought to be the expanding 
supernova shells of the progenitor star of the black hole in SS~433. 
The source has been searched for by the HEGRA, MAGIC, and CANGAROO-II Cherenkov telescopes \citep{HegraSS433, CangarooSS433}, 
resulting in upper limits for the gamma-ray emission from both the inner system and the different interaction regions with the W50 nebula. 
Strong gamma-ray absorption is expected from this system, due to the periodic eclipses through the companion star as well as attenuation due to the precession of the accretion disk envelope \citep{Reynoso08}.
As in the case of LS I+61 303, the presence of a significant
amount of target material for cosmic ray interactions as well as the possibility
of a higher energy emission than what is inferred
from gamma-ray observations due to absorption, makes this an interesting
candidate for a neutrino source. SS~433 has also been tested for possible periodic neutrino emission in \citep{MikePaper}.

\subsection{The Cygnus Region}
\label{method_cygnus_region} 

The Cygnus region is roughly located within Galactic longitudes $70^{\circ} < \ell < 90^{\circ}$ 
and latitudes $-4^{\circ} < b < 8^{\circ}$, where our line of sight is directed nearly along the local spiral arm of the Galaxy \citep{2008hsf1.book...36R}.
At a distance of approximately 5~kpc our line of sight has left the local arm and crosses the 
Perseus arm, and even the outer arm further away ($\sim$10~kpc).
Here many different sources are located at various distances superimposed in a relatively small area in the sky,
resulting in a complex region which harbors some of the closest and most massive regions of star formation in the Galaxy. 

The vast majority of the molecular gas detected in the Cygnus region is concentrated on the local arm \citep{Schneider06}, at distances between 1-3~kpc. One of the most massive giant molecular cloud complexes
in the Galaxy resides within this region, at a distance of $\approx$1.7~kpc. It is thought to be the birth place of the massive Cyg OB2 association, and probably also Cyg OB9 and Cyg OB1, as well as of a number
of less massive star clusters with young stars or ongoing star formation \citep{LeDigou08}. The strong stellar winds and radiation pressure of the massive stars in the Cygnus region have strongly influenced
the spatial distribution of the molecular gas in the region, displacing and compressing the gas forming filamentary structures and dense clumps which surround the less dense environment of the cluster,
in which the gas has been evacuated. If high-energy particles are generated within the stellar associations, they can interact with the ultraviolet radiation fields producing TeV gamma-rays
through the Inverse Compton and p$\gamma$ processes. However, protons and nuclei can travel longer distances than electrons, and they may also leave the photon dominated regions around
the massive star clusters and interact with the nearby molecular clouds. The resulting neutrino flux map would then reflect the complicated distribution of the gas in the region.
It is also worth noting that the injection of cosmic-rays may take place at 
multiple locations due to presence of several particle accelerators inside the Cygnus region. 
The existence of these accelerators is confirmed by the observation of strong TeV gamma-ray emission throughout an area of approximately $10^\circ \times 10^\circ$ \citep{MgroCygnus,AliuCygnus}.
The potential for IceCube to observe neutrinos from this region has been discussed in \citep{AnchCygnus, BeacomCygnus, KappesCyg}, based on the measured TeV gamma-ray flux. 

Due to the complexity of the possible spatial distribution of events within the Cygnus region, we have applied an analysis of the spatial correlations between neutrino events to search
for an astrophysical neutrino signal in an extended region. If the intensity fluctuations of a possible neutrino signal throughout the region follow a certain correlation structure,
this may show up as a significant deviation from the random distribution of atmospheric neutrino events. 

In the IceCube analysis, we use the two-point correlation function formalism introduced by Peebles \citep{Peebles} and co-workers \citep{PeeblesGroth75, FryPeebles78} to study the large-scale matter distribution
in the Universe \citep{madox96}. In particle astrophysics, the correlation function has been used to search for anisotropies in the spatial distribution of cosmic-rays \citep{westerhoff04, chad04} and
neutrinos \citep{abbasi09AMANDA}. 

Here we use the approach to search for neutrinos inside an area of $11^\circ \times 7^\circ$ centered on the most active part of the Cygnus complex in TeV gamma-rays.
We define our test statistic in terms of a clustering function, $\Phi(\Theta)$, which is the excess or deficit in the number of event pairs within a certain distance
with respect to the background-only hypothesis (similar to a cumulative correlation function, see e.g. \citep{Peebles} and references therein,
\citep{CygnusKerscher, CygnusLandy} for estimators of the two-point correlation function based on pair counting): 

\begin{equation}
\Phi(\Theta) = \frac{\int_0^{\Theta}DD(\Theta) d\Theta}{\int_0^{\Theta} RR(\Theta) d\Theta}
\label{eq:clusfunc}
\end{equation}
where $\Theta$ is the distance between two events, and $DD(\Theta) = \displaystyle\sum_{ij}DD_{ij}$, $RR(\Theta) = \displaystyle\sum_{ij}RR_{ij}$, where the sum is over all non-repeated pairs in, respectively, the real data sample and in a sample randomized in azimuth (representative of a pure background case). In our case, $DD_{ij} (RR_{ij}) = 1$ only if either the event $\it{i}$ or the event $\it{j}$, or both, are within the region under study, and it is equal to zero otherwise. It is worth noting that with this definition we measure both the intensity of the process that generated the observed neutrino event pattern as well as its correlation structure.

\subsection{Cygnus X-3 Flares}
\begin{table*}[t!]
\begin{center}

\caption{Criteria followed for the selection of Cyg~X-3 flaring
periods.}
\small{
\begin{tabular}{ccrr}
\hline
Wavelength & Telescope & START & STOP \\
\hline 
Radio  & AMI radio telescope \citep{zwa08} & $S_{15\un{GHz}} > 1\un{Jy}^a$ & $S_{15\un{GHz}} < 1\un{Jy}^a$ \\
\multirow{2}{*}{X-ray} & RXTE/ASM \citep{lev96} \& & \multirow{2}{*}{$S_{\mathrm{fit}} > 1\un{Jy}^b$} &
\multirow{2}{*}{$S_{\mathrm{fit}} < 1\un{Jy}^b$} \\  & Swift/BAT \citep{bar05} & \\
\hline \\
\multicolumn{4}{l}{$^a$ $S_{15\un{GHz}}=$ Measured radio flux density at 15 GHz.}\\
\multicolumn{4}{l}{$^b$ $S_\mathrm{fit} = \kappa \,\exp{-(t-t_0)^2/2\sigma^2}$, average of Gaussians fitted to 28 radio
flares, $\kappa = 4.81\un{Jy}$, $\sigma = 1.16\un{d}$. }\\
\multicolumn{4}{l}{One average radio flare is centered on each X-ray state with hardness $>$ 0.001 following a state}\\
\multicolumn{4}{l}{ with hardness $<$ 0.001 within 10 days since those are candidate radio flare events.}\\
\end{tabular}}
\label{CygX3_flares}
\end{center}
\end{table*}

\begin{figure}[t!]
\begin{center}
\includegraphics[width=5.0in]{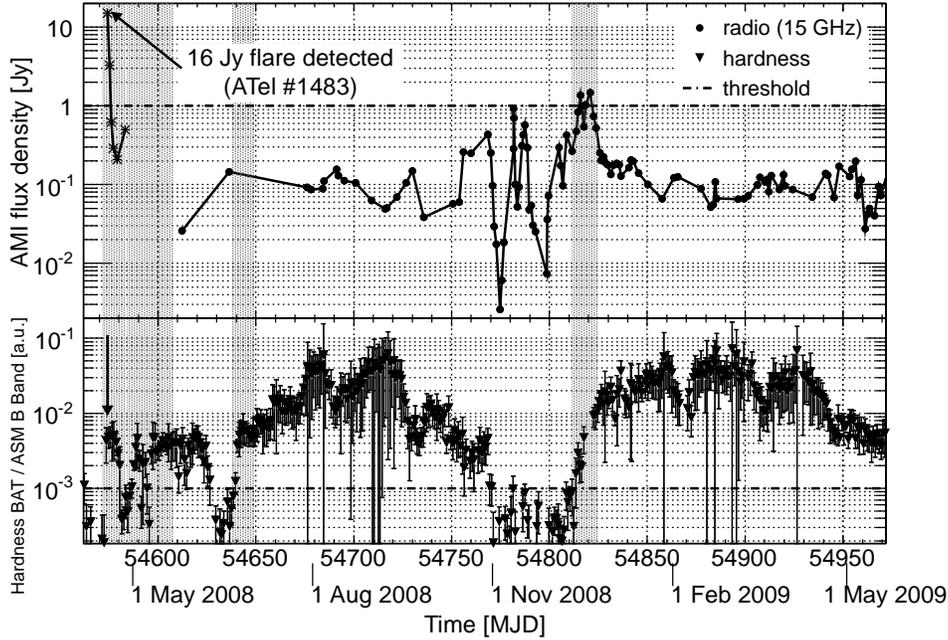}
\caption{The radio light curve and hardness of Cyg~X-3 with the four time windows of the analysis (gray shading). }
\label{fig_windows}
\end{center}
\end{figure}

The high-energy sky presents strong variability in time. If neutrino emission from
a particular object is expected to vary with time and if it is in coincidence with 
electromagnetic emission, it
is advantageous to include the time information in the data analysis.
One such case is the microquasar \objectname{Cygnus X-3} (Cyg~X-3) at a distance of $\approx 9$\,kpc. 
Coinciding with the period of data taking considered in this analysis,
first observations of production of high-energy photons within the system of Cyg X-3 were published by the Fermi
\citep{abd09} and AGILE \citep{tav09} satellite missions independently. The reported gamma-ray
fluxes are in the energy band between 100~MeV and 100~GeV (Fermi), following a power-law
of spectral index $-2.70\pm0.25$, and between 100~MeV and 10~GeV (AGILE) with spectral index
$-1.8\pm0.2$. However, gamma-rays from Cyg X-3 are only observed during certain periods
of time, probably correlated with strong radio outbursts and certain X-ray emission states of
the system \citep{abd09,kol10}. No TeV gamma-rays from Cyg X-3 have been detected by Cherenkov
telescopes like MAGIC and VERITAS so far, a possible explanation being strong absorption of
high-energy photons or limited observation time. Upper limits on the integrated gamma-ray flux above 250 GeV are at the level of $2.2 \times 10^{-12}~\rm{cm^{-2} s^{-1}}$ \citep{MagicCygX3}.
Assuming a hadronic origin of at least part of the gamma-rays, the high densities
in the system of Cyg X-3 could provide an environment for copious neutrino production at TeV
energies, detectable by neutrino telescopes such as IceCube \citep{abd09,bed05}.
A previous search for periodic neutrino emission was reported in \citep{MikePaper}. 

Here we perform a search for neutrino emission during, or close to, the observed flaring activity in Cyg X-3. The phenomenology of emission
states of Cyg X-3 has been studied carefully using radio and X-ray observations \citep{szo08,tud07,kol10}. 
With radio and X-ray data, there are
two ways to identify active periods of Cyg X-3 associated with jet ejection: the observation of
a radio flux above 1 Jy, following \citep{szo08}, and the observation of hyper-soft X-ray states
and subsequent hardening of the X-ray spectrum, following \citep{kol10}. The identification
of potential flaring periods of Cyg X-3 is thus split into a radio and an X-ray part, as in Table~\ref{CygX3_flares}. 
For reconstruction of radio flares from X-ray
data, an average radio flare was obtained from Gaussians fitted to 28
flares seen to rise above 1~Jy in radio data. One average radio flare was
put at each time of potential radio flaring, as seen in X-ray data (i.e.
spectral hardness > 0.001 within 10 days after a state with hardness
< 0.001) to obtain a pseudo radio flux density $S_\mathrm{fit}$.
The resulting search windows from all selections can
be overlapping and are combined with a logical OR operation, resulting in the final time windows.
The utilized radio data were taken with the Arc\-mi\-nu\-te Microkelvin Imager (AMI) radio
telescope \citep{zwa08,poo97} at a frequency of 15~GHz between May 2008
and May 2009 in irregular intervals. X-ray data were obtained from the Rossi X-ray Timing
Explorer/All Sky Monitor (RXTE/ASM) \citep{lev96}, using the B band between
3 and 5~keV, and from the Burst Alert Telescope on board the Swift satellite (Swift/BAT)
\citep{bar05} sensitive between 15 - 50~keV. 
\footnote[1]{AMI data from Pooley,G., \url{http://www.mrao.cam.ac.uk/\~guy/cx3/data/} (2010-01-14). 
RXTE/ASM data from Bradt, H., Chakrabarty, D., Cui, W. et al., \url{http://xte.mit.edu/ASM\_lc.html} (2010-03-11). 
Swift/BAT data from Krimm, H., \url{http://swift.gsfc.nasa.gov/docs/swift/results/transients/CygX-3/}
(2010-06-04).}
The ratio of BAT (hard X-ray) to ASM B (soft X-ray) counts provides the spectral
hardness parameter. Gamma-ray data from Fermi or AGILE are not explicitly taken into consideration
in this analysis.
Applying the selection criteria described in Table~\ref{CygX3_flares} to the radio and X-ray data from the time
when IceCube 40-strings was operating (between MJD 54560 and MJD 54989) results in the four time windows indicated in Table~\ref{windows}
and Figure~\ref{fig_windows}. Even though there are no AMI data from the first 2.5 months of
this period, an ATel \citep{tru08} was issued for a strong radio flare around MJD 54574
that is consistent with the time windows selected from X-ray data.

\begin{table}[t!]
\begin{center}
\caption{\footnotesize Search windows extracted from radio and X-ray data for the neutrino search from Cyg~X-3 direction.}
\begin{tabular}{cccc}
\tableline\tableline
\multirow{2}{*}{Window} & Start & Stop  & Duration \\
               & (MJD) & (MJD) & (days) \\
 \tableline
    1 & 54571.4     & 54582.5    & 11.1 \\
    2 & 54584.5     & 54607.4    & 22.9 \\
    3 & 54637.6     & 54649.5    & 11.9 \\
    4 & 54811.5     & 54824.6    & 13.1 \\
\tableline
\end{tabular}
\label{windows}
\end{center}
\end{table}

The data have been analyzed with a maximum likelihood test using a time-dependent
version of the unbinned likelihood ratio method \citep{bra10}. The search time windows
are incorporated into the signal probability density function (p.d.f.) of the likelihood
function as normalized Gaussians with mean located at the window center and FWHM equal to the window
duration. During the maximization, the windows were allowed to be shifted up to
20 days to earlier or later times. This allows us to find neutrino emission that comes before or
after a radio flare. The value of 20 days is motivated by the hypothesis of emission during
the radio quenched state, which can happen up to $\sim20$ days before the onset of a major radio 
flare \citep{kol10,tru07}. As in the other analyses presented in this work, no energy estimator is used in the likelihood.
In the search for neutrinos from the microquasar Cygnus X-3, five searches are performed in total:
one with each of the four windows as a hypothesis of a neutrino signal light curve, shifting each window individually, 
and one search using all four windows, shifting the windows simultaneously. Doing this analysis,
only about $50\%$ of the discovery flux of a time-integrated search (that uses no information about
the activity of Cyg X-3) is needed for a $5\sigma$ discovery, assuming the neutrino emission
happens during the windows or within $\pm20$ days. The discovery flux with $50\%$ detection
probability is $\rm{dN/dE} \approx 10^{-10}~\rm{TeV^{-1} cm^{-2} s^{-1}}$ for
an E$^{-3}$ spectrum and $\rm{dN/dE} \approx 1.2 \times 10^{-11}~\rm{TeV^{-1} cm^{-2} s^{-1}}$ for
an E$^{-2}$ spectrum.

\section{Neutrino Samples}
\label{Sec4}

We discuss here the selection of the sample of neutrino candidates for 
each of the two considered detector configurations. While the event
selection is optimized for each of these two configurations
separately, the two analyses are both based on the selection of events on
the basis of track reconstructions. We therefore first discuss the
concepts that are common to the two event selections: the track
reconstruction and its application in on- and off-line event selection as
well as a special filtering of the AMANDA data.

\subsection{Track Reconstruction and Event Filtering}
\begin{figure}[t!]
\begin{center}
\includegraphics[width=0.9\textwidth]{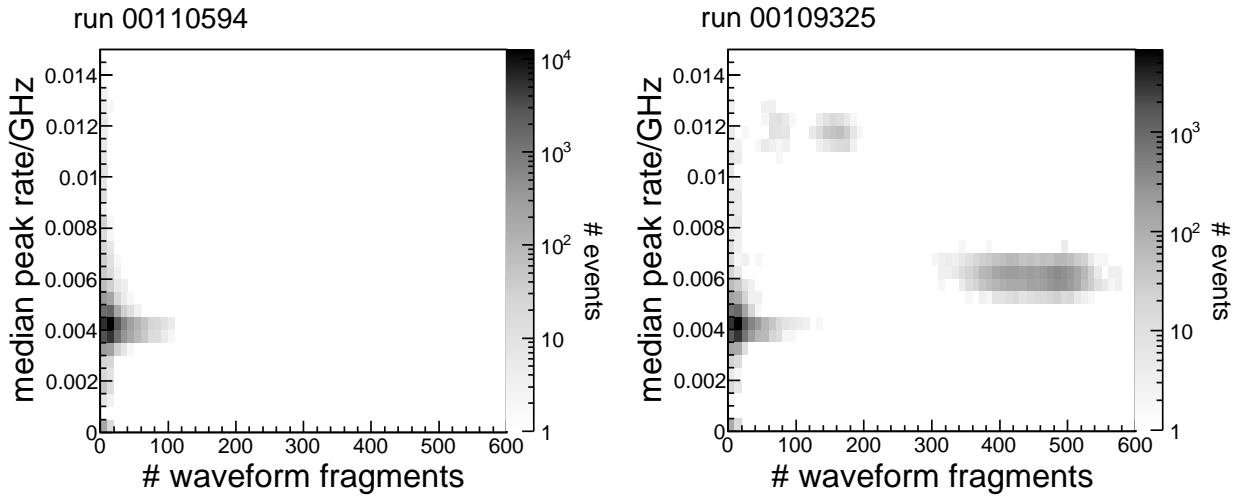}
\caption{Distributions of the parameters used to identify non-particle
induced events in AMANDA. Shown are the median peak rate per waveform 
versus the number of waveform fragments (see text) on strings 5-10. 
The right plot represents a run with stronger pickup of electromagnetic
noise (109325) while the left plot is obtained from a normal run
(110594).
Events which have more than 20 waveform fragments on strings 5-10 and a median
peak rate above 0.005~GHz are rejected as non-particle induced events.}
\label{FlareCheck}
\end{center}
\end{figure}

At energies above $\approx$200~GeV, the identification of
neutrino candidate events in the IceCube and AMANDA data is based on the
selection of well-reconstructed up-going events since the dominant 
background are muons from cosmic ray air-showers, which reach the detector in the downward direction but are filtered out by the Earth in the upward direction, leaving only neutrinos. 

The IceCube and AMANDA data are divided into two streams which are treated in a similar way.
The first stream are events that trigger AMANDA. As explained above, these events are then complemented with the data
collected in IceCube and for this reason they are considered as combined
events (C-events). The second stream are those events which trigger
only IceCube (ICO-events), either because the number of hits in AMANDA is below
the trigger threshold or because AMANDA was not active at the time the
event was recorded.

Initial fast track reconstructions and event selections are applied
on-line at the South Pole since the available bandwidth for satellite
transfer of data is limited. For ICO-events, a straight track is fit to the
data by minimizing the distance between the track and the hits (linefit)
\citep{AMANDARECOPAPER}. For C-events, a pattern recognition algorithm
(called JAMS) has been used \citep{Markus_thesis}. The on-line filters are based on these
reconstructions and the relevant events for the presented analysis are selected by
requiring that the first-guess reconstruction is not down-going. At this
level, the data are still dominated by atmospheric 
muons that are mis-reconstructed as up-going. 
In particular, coincidences between multiple muons from different 
cosmic-ray air showers can mimic up-going event topologies.

After transfer to the north, more sophisticated reconstructions are
applied to improve the angular resolution and to provide quality
parameters for the rejection of the down-going atmospheric muon
background. These reconstructions are maximum likelihood fits which are
based on the probability density function for the arrival time of a
photon given the track hypothesis \citep{2007APh....28..456V}. Two
likelihood reconstructions have been studied: one based only on the first
photon in each optical module (SPE) and a more complete one which includes
the possible presence of multiple photo electrons (MPE). For more
details about the SPE and MPE reconstructions, we refer to
\citep{AMANDARECOPAPER}. A list of variables that are indicative of the quality of the
reconstructed event is given in \citep{2011ApJ...732...18A}.
In addition to the track reconstruction itself, an estimate of the angular
uncertainty of the track reconstruction is obtained for each individual
event by the evaluation of the likelihood function near the maximum. This
method is described in \citep{2006APh....25..220N}. The estimate of the angular uncertainty
is used for the event selection as well as in the maximum likelihood ratio
test.

In the IC40+A analysis, we also make use of an
energy reconstruction which is based on the characterization of the energy
loss along the particle track. At energies above a few hundred GeV, the
energy loss of a muon in the ice is proportional to its
energy. The energy reconstruction used in this work is presented in
\citep{2008ICRC....5.1275Z}.

\subsection{AMANDA data}
\label{Sec4.1}

In contrast to IceCube, the waveforms collected with the optical modules of
AMANDA are not digitized directly in the optical modules but are
transferred to the surface as analog signals. This introduces two undesired effects in the data. 
The first one is crosstalk between different cables. 
Large pulses in one optical module can cause a
detectable signal within the electronics of cables connected to other optical modules.

The second issue concerns the pickup of electromagnetic noise. The PMT signals
have to be transferred over a distance between 1.5~km and 2.0~km to the surface. The
cables needed for this task are vulnerable to pick up electromagnetic
noise as they act as electromagnetic antennae.

Methods dedicated to the identification of non-particle induced signals based on
the waveforms have been developed and result in an efficient separation
from particle-induced signals. The integral over the entire collected 
waveform pulse is used in order to remove crosstalk pulses. Since
crosstalk pulses do not originate from a charge deposit in the PMT, they
consist of (positive and negative) fluctuations around the baseline with
the total integral close to zero.
Waveforms from a particle-induced signal in the PMT have a characteristic
width, which is wider at the surface due to dispersion in the cables. 
For AMANDA strings 5-10, this is typically 250-300~ns for single
photoelectrons (SPE). Waveforms from multiple photoelectrons (MPE)
result from the (linear) overlay of many SPE waveforms and typically are
wider than those. In contrast, noise-induced waveforms are often
very spiky, i.e.\ they have many peaks within the width corresponding to a
typical SPE pulse. This feature of non-particle-induced waveforms
has been used to remove noisy events from the data: events which simultaneously
have a median peak rate in the waveforms recorded in AMANDA strings 5-10
which is incompatible with a PMT signal (above 5~MHz) and a high number (more than 20) of waveforms in
these strings, are considered non-particle induced and are
removed from the data set. Figure~\ref{FlareCheck}
illustrates this cut. Both analyses presented in this paper first apply cross-talk 
cleaning and the above technique for rejection of non-particle induced events to 
the AMANDA data, before further event selections are made.

\subsection{IceCube 22-strings and AMANDA}

Data have been collected from May 31, 2007 until April 4, 2008 when IceCube was 
operating in a 22-string configuration. The lifetime of the IC22+A
run is 276~days, including 143~days of AMANDA operating in stable 
mode. The unusually long downtime of AMANDA during this period was 
caused by various hardware failures during May 2007 (trigger system) and
during August 2007 (high voltage supply system). In this section, the
event selection from the trigger level up to the final analysis level is
described and the characteristics of the combined neutrino sample
are highlighted.

\subsubsection{Trigger and on-line Filter}

As explained above, IceCube and AMANDA are triggered separately in the combined detector mode.
The trigger rate of IC22 is 550~Hz, while AMANDA triggers at
200~Hz. Seasonal variations affect the trigger rate by about $10\%$. 
The overall trigger rate of the combined
IC22+A detector after correction for overlaps
between the two triggers is 640~Hz. At trigger level, the data are
strongly dominated by down going atmospheric muons induced by cosmic-ray
air showers outnumbering atmospheric neutrinos by a factor of about $10^6$.
The on-line filter reduces the data volume for satellite transfer and has
a passing rate for reconstructed up-going or horizontal events of 22~Hz
for ICO-events and 8~Hz for C-events, producing a total event rate of
30~Hz.
\begin{figure}[t!]
\begin{center}
\includegraphics[width=3.5in]{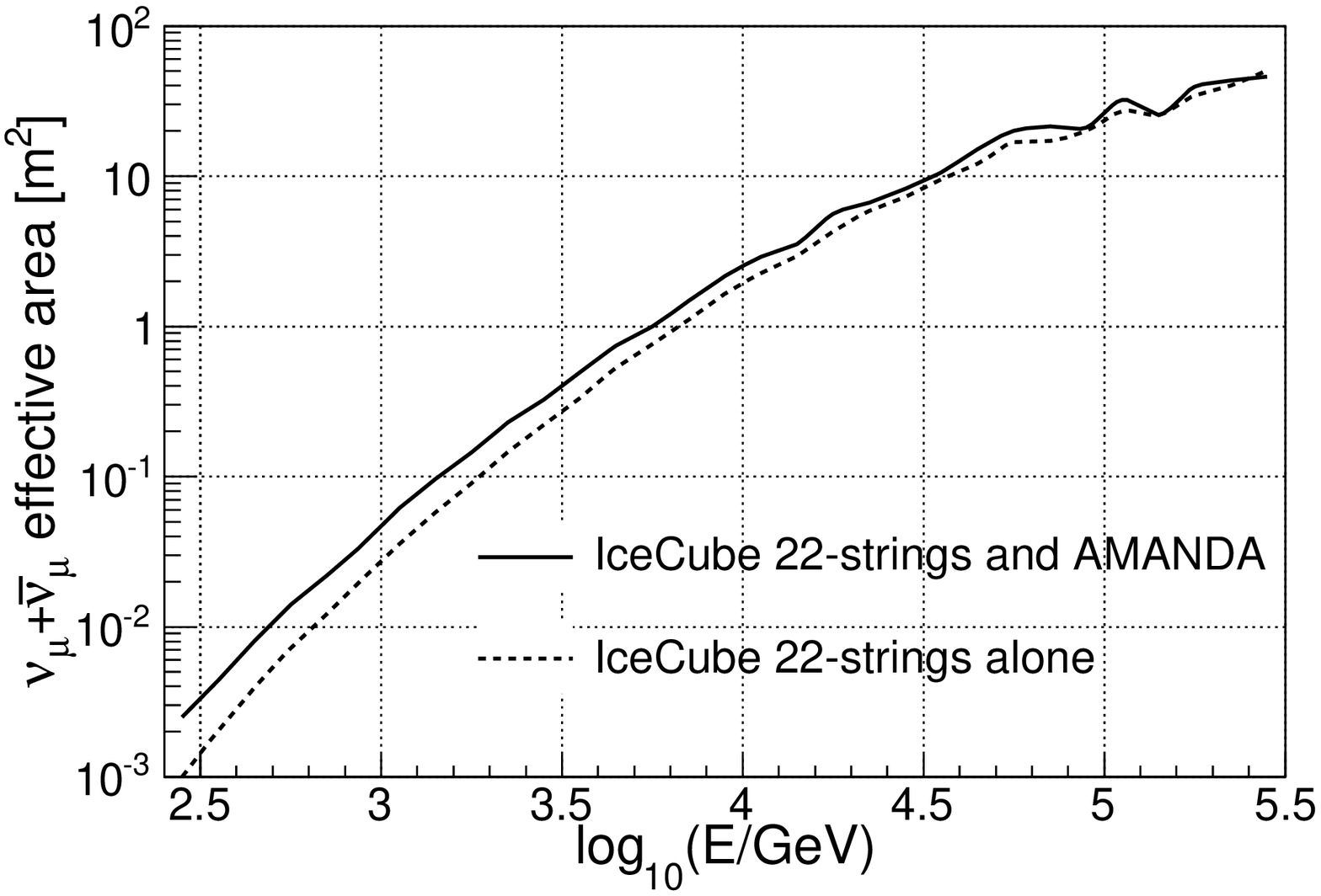}
\includegraphics[width=3.5in]{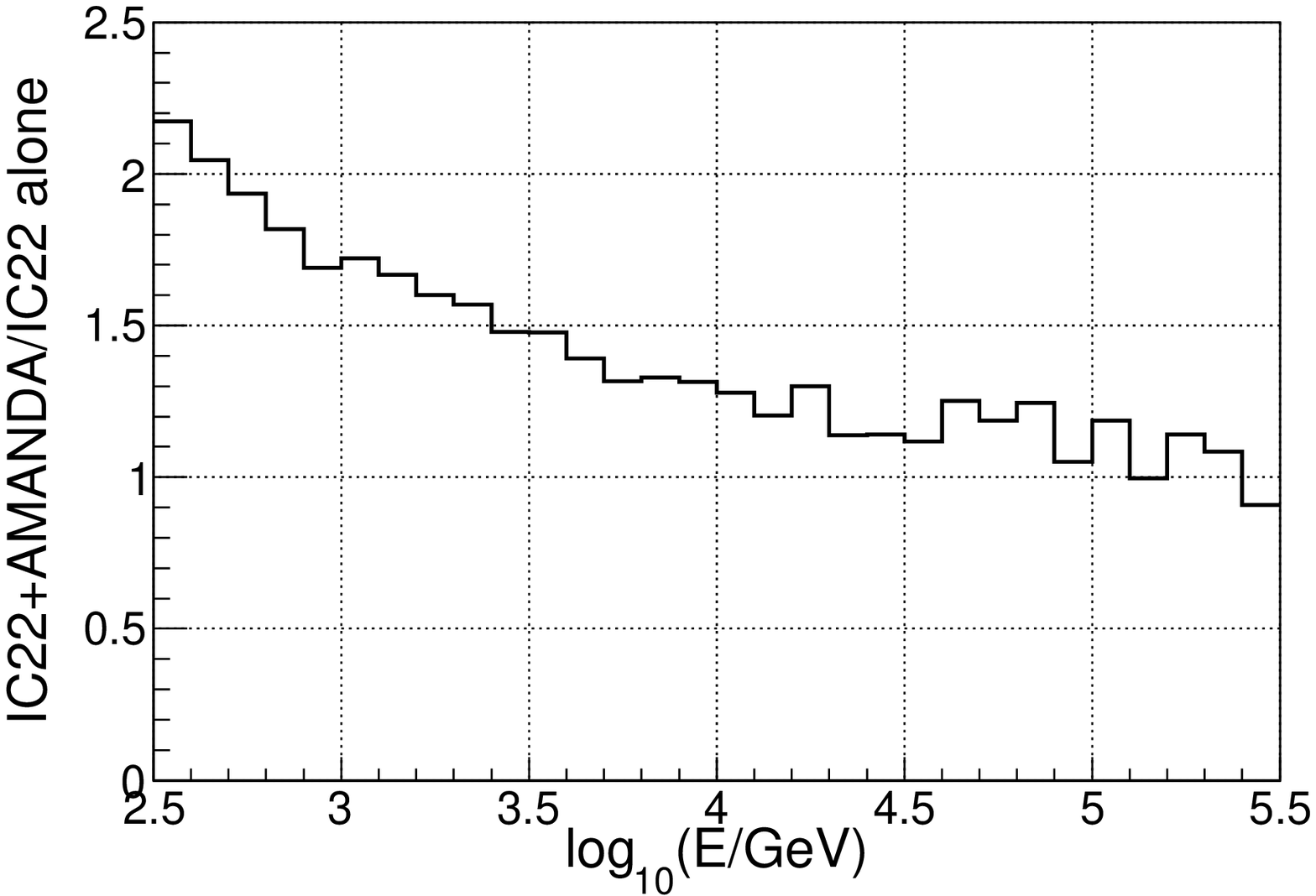}
\caption{Comparison between the effective areas obtained in this work for IC22 with and without AMANDA. The effective area of the combined detector at 1~TeV is a factor $\approx$1.6 larger than without AMANDA.
The last bins of the ratio plot are affected by limited statistics.}
\label{a_eff_IC22AMA}
\end{center}
\end{figure}

\subsubsection{Neutrino Sample}
After the rejection of down-going reconstructed events based on first guess
reconstructions, the data are still dominated by mis-reconstructed
atmospheric muons and further event selections are needed in order to
arrive at a sample of events dominated by atmospheric neutrinos. A typical
quality parameter used in this analysis is the number of 
un-scattered photons, so-called "direct hits," which are characterized by a small time residual with 
respect to the expectation of the geometry of the emitted Cherenkov cone.

In order to optimize the retention of lower energy events we have 
employed a multivariate approach. In this approach, a signal likelihood is 
defined as the product of the signal likelihood from the considered
variables and is compared to the likelihood of the background hypothesis.
According to the Neyman-Pearson lemma \citep{NP_lemma}, this
criterion leads to the best possible discrimination power for the given
set of variables if there are no correlations between them. For correlated
variables, as in our case, this criterion turns out to still be 
powerful. We use experimental data dominated by atmospheric muons 
to describe the background, and simulated neutrinos weighted to an 
atmospheric neutrino spectrum in order to model the
signal. C-events and ICO-events are treated independently here. For
ICO-events, further cuts are used to reject coincident air shower muons.
These cuts are based on the smoothness of the distribution of the hits
along the track and on reconstructions performed on subsets of hits.
For C-events, the rate of coincident muons is significantly lower due to
the smaller size of AMANDA. While a tighter time-window cleaning helped to
further reject the coincident muons in the combined events stream, no
dedicated cuts/reconstructions have been used to remove these.

The final cuts were optimized for the best discovery potential, and
finally tracks are selected if the angular resolution estimator returned a
value lower than $4^\circ$. A harder cut on this parameter did not lead to an improved 
discovery potential.

\begin{figure}[t!]
\begin{center}
\includegraphics[width=4.0in]{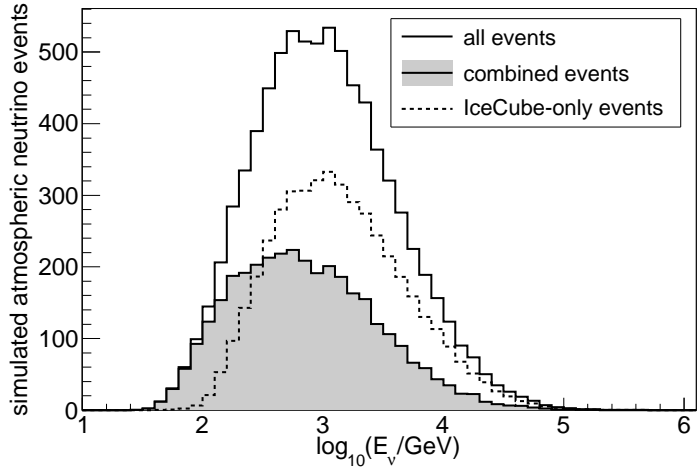}
\caption{Energy distribution for simulated atmospheric neutrinos at the
final selection level of the Galactic point source analysis with IC22+A
normalized to the lifetime of the IC22 data.}
\label{atm_E_IC22_AMA}
\end{center}
\end{figure}

The resulting neutrino sample contains about 1.8 times more events
than the search presented in \citep{ic22ps}. In total, 8727 events
are selected, of which 3430 are C-events. These event numbers correspond
to a data rate of $4.7\cdot10^{-4}$ Hz for IC22+A 
and to $2.4\cdot10^{-4}$ Hz in IC22 only mode. 
The effective area of the IC22+A analysis
is shown in
Figure~\ref{a_eff_IC22AMA}. Below energies of a few TeV, there is a strong
improvement when including AMANDA, with respect 
to the performance of a low-energy optimized analysis using IceCube only. The 
energy distribution of atmospheric neutrinos for this selection according
to simulations is shown in Figure~\ref{atm_E_IC22_AMA}.

The estimated angular resolution is given by a median of $1.9^\circ$ for
atmospheric neutrinos and $1.7^\circ$ for a Crab-like spectrum. About
$10\%$ of the final events are expected to be mis-reconstructed muon
background. 

The minimum detectable flux is more than an order of magnitude above the
neutrino emission expected from the Crab assuming that the H.E.S.S.
observations are consistent with a model of $pp$ interactions. While with
this expectation, a positive detection is unlikely but not excluded as the
photon flux may be absorbed, the analysis is valuable as a starting point
for future improvements with the full IceCube detector and the DeepCore
sub-detector. Using an E$^{-3}$ spectrum for the comparison, the discovery flux
in this analysis is between 15\% and 33\% lower than the one in \citep{ic22ps}, depending on the declination. This clearly demonstrates the improvement obtained by the combined use of a low-energy core and the optimization of the event selection for softer spectra. An additional improvement is obtained by the lower number of trials in the Galactic Plane scan versus the scan of a hemisphere or the whole sky but is not quantified here.

\subsection{IceCube 40-strings and AMANDA}

Following the explorative analysis strategy developed for the IC22+A, a similar analysis has been conducted on
the larger data sample collected with the combined IC40+A
detector from April 5, 2008 to May 20, 2009. Both
parts of the combined IceCube-AMANDA detector operated very stably during
this time period. For IC40 about 375~days of data were collected and used in
this analysis and for AMANDA about 306~days.  
The main causes for
downtime were scheduled operations in the course of the integration of new
strings into the detector. Moreover, the decommissioning of AMANDA began a few weeks
before the completion of the IC40 run. The event selection
is in many aspects similar to the one applied to the IC22+A
data as the targeted energy range is the same as
well as the physics driving the analysis. Again, different cut criteria
are developed for C- and ICO-events.

\subsubsection{Trigger and on-line Filter}

For what concerns this analysis, the trigger logic was kept identical 
to the one in the previous season. The on-line filter selected about 20~Hz of track-like ICO-events 
and 3~Hz of up-going C-events.

\subsubsection{Neutrino Sample}

\begin{figure}[t!]
\begin{center}
\includegraphics[width=4.0in]{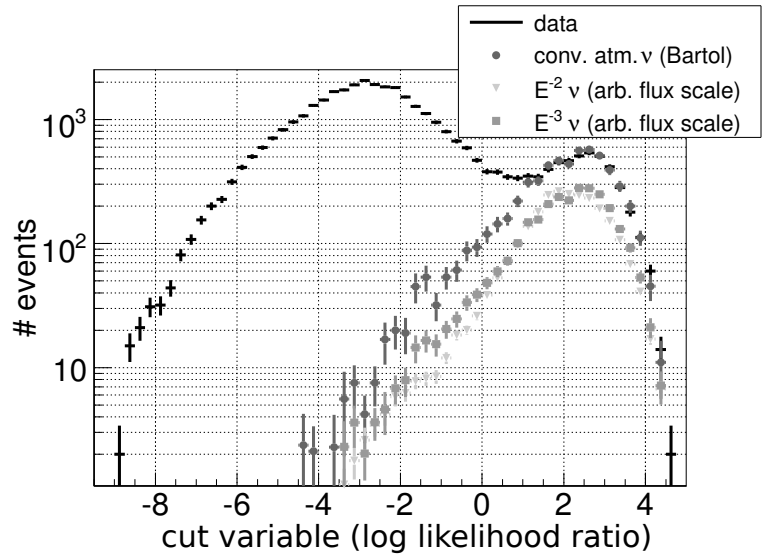}
\caption{Distribution of the main cut variable for IceCube events in the IceCube 40-strings and AMANDA analysis.
Data are shown along with simulated atmospheric neutrinos \citep{Bartol04} and two different simulated neutrino signals of arbitrary flux scale.
No normalization is applied.
A cut at a value of 1.0 is applied to the data to select a neutrino sample.}

\label{IC40AMA_NP}
\end{center}
\end{figure}

Similar to the previous combined analysis, ICO-events are selected by a
series of one-dimensional cuts on event quality parameters and combined
with a multivariate classification based on the Neyman-Pearson lemma
\citep{NP_lemma}. The probability density functions for five quality
parameters are generated from down going atmospheric muon-dominated data as background and from
up going atmospheric neutrino simulation as signal, and combined in the cut. (The fraction of 
atmospheric neutrino events in the data is still only about 4\% and the data 
can therefore be regarded as dominated by background atmospheric muons.) 
The five quality parameters used in this analysis are: the quality parameter 
of the likelihood reconstruction, an estimate of the angular uncertainty
of the likelihood reconstruction obtained by the evaluation of the likelihood space around
the maximum, and three variables which describe the amount and distribution
of unscattered light in the event. A PMT pulse from un-scattered light is characterized by
a small time residual with respect to the expectation from the geometry of the Cherenkov cone.
The number of PMT pulses with time residuals between -15~ns and 75~ns, the maximum distance between their projections
on the reconstructed track and the smoothness of their distribution along the track of the particle
are used in the event selection. More information about these variables is 
reported in \citep{2011ApJ...732...18A}. The distribution of the resulting cut variable is shown in
Figure~\ref{IC40AMA_NP} for data and for atmospheric neutrino simulation
as well as for two example signal neutrino spectra. An optimization of the
discovery potential for a soft E$^{-3}$ spectrum results in an optimal cut
value of 1.0.

\begin{figure}[t]
\begin{center}
\includegraphics[width=4.0in]{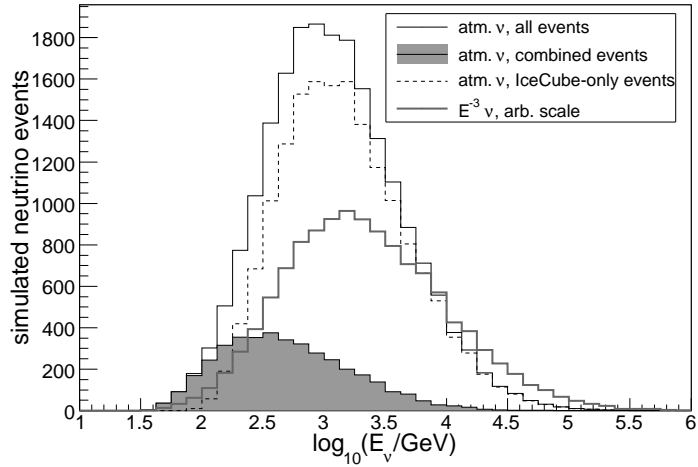}
\caption{Energy distribution of the final neutrino sample obtained in the IceCube-40 strings and AMANDA data sample.
This comparison is based on simulated neutrinos following the Bartol model \citep{Bartol04}.}
\label{IC40AMA_MCEN}
\end{center}
\end{figure}

C-events are first cleaned as described in section \ref{Sec4.1}. 
Subsequently, a series of one-dimensional cuts is applied. This series of cuts has been
challenged versus other more sophisticated cut strategies and
demonstrated to perform well enough in the separation of the atmospheric
muon background and the neutrino signal.

While the analysis is optimized for soft spectra such as E$^{-3}$ or the
Crab-spectrum, it is nevertheless desirable to retain 
a good efficiency for very high-energy neutrino events as well. With a cut
optimized on a very soft spectrum however, the retention of high-energy
events is not necessarily optimal as these deposit much more light in the
detector and may thus have event topologies that are not caught in the
low-energy event selection. To remedy this, additional criteria are included
in the event selection if the signature is likely to be induced by
a very high-energy neutrino. These additional events are selected in both
streams with a series of one-dimensional cuts based among others on their
reconstructed energy.

The IC40+A analysis also uses maximum likelihood track reconstructions.
For low energetic combined IceCube-AMANDA events, a single photo
electron (SPE) pdf is used and for all other events, a multiple photoelectron (MPE) pdf is used.

\begin{figure}[b] 
 \begin{center} 
\includegraphics[width=0.4\textwidth]{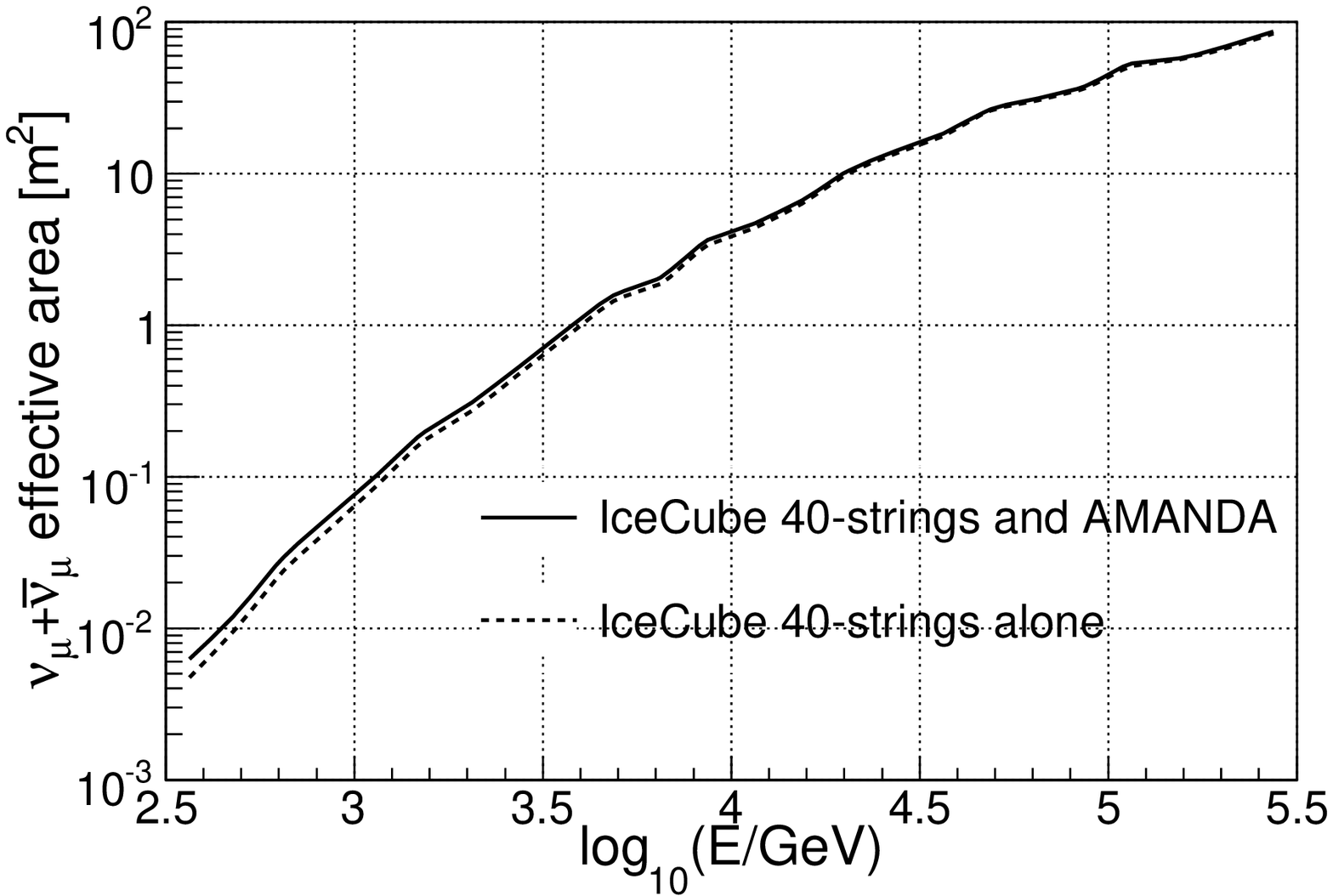} 
\includegraphics[width=0.4\textwidth]{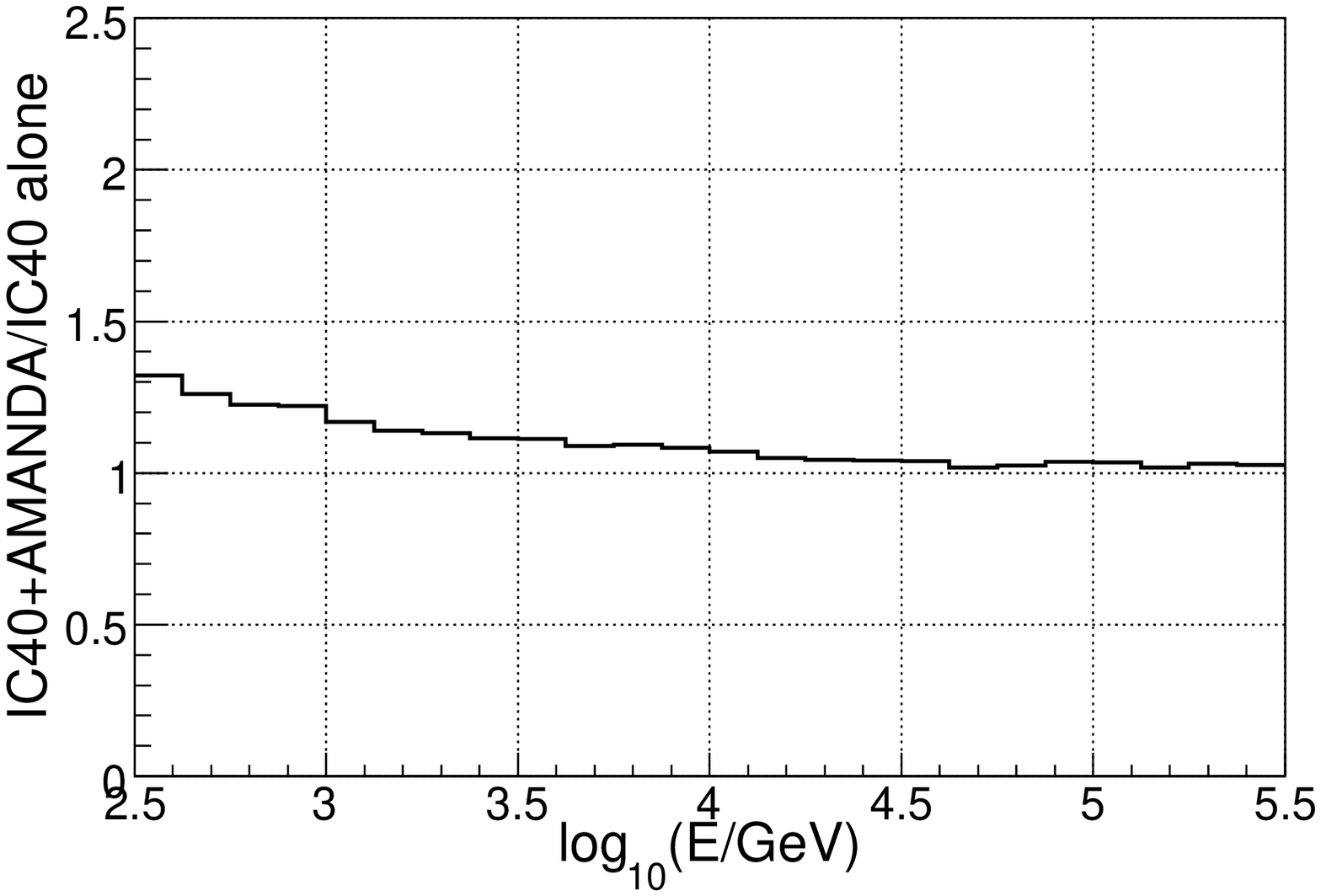} 
\caption{Comparison between the effective areas obtained in this work for IceCube-40 strings with and without AMANDA. Depending on the declination angle the increase at 1~TeV due to AMANDA is between
10 and 20\%. Moreover, the effective area achieved with IC40+AMANDA at 1~TeV is a factor 1.4-2.0 larger than that in the IceCube-40-only analysis \citep{2011ApJ...732...18A},
due both to the inclusion of AMANDA and to an event selection optimized for lower energies.} 
 \label{IC40_EFF_AREA_COMPARISON} 
 \end{center} 
\end{figure}

The total number of selected neutrino candidates is 19,797 in the entire lifetime of 375~days. The purity of
the atmospheric neutrino sample is estimated to be 97-98\%.
Of the selected neutrino candidates, 81.3\% are
ICO-events selected with the multivariate Neyman-Pearson likelihood ratio
cut. 2.4\% are additional ICO-events with high estimated energies. The
remaining 16.3\% events are C-events. Despite the larger lifetime of
AMANDA in 2008/2009 with respect to the previous year, the fraction of
C-events in this analysis is smaller than in the previous one. This is
partially due to the larger size of the IceCube detector but also to the
higher purity of the IC40+A sample. The resulting energy
distribution as derived from atmospheric neutrino simulation is shown in
Figure~\ref{IC40AMA_MCEN}. The selected C-events peak at lower energies than the ICO-events. Also the
effective area, reported in Figure~\ref{IC40_EFF_AREA_COMPARISON}, shows the
effect of AMANDA at the lower energies. With the larger size of the detector, the effective area is improved significantly
with respect to the IC22 analysis.

\begin{figure}[t!]
\begin{center}
\includegraphics[width=4.0in]{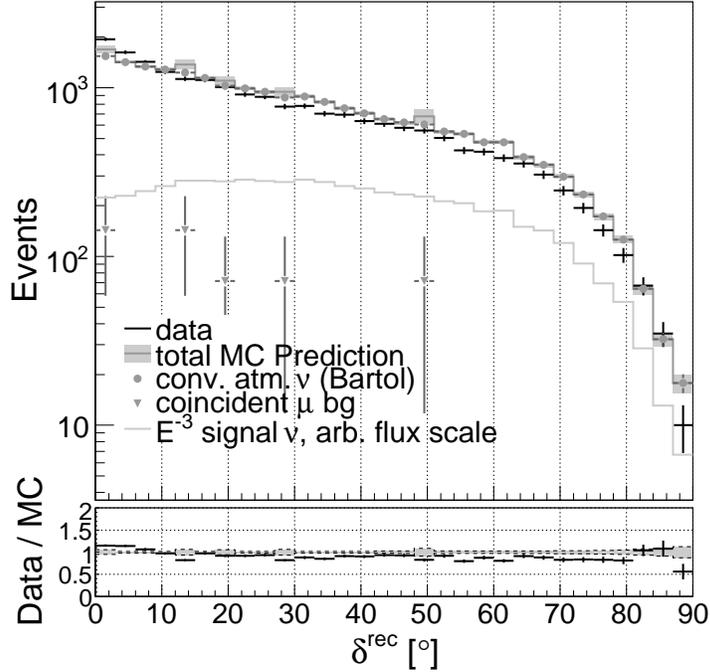}
\caption{
Comparison of experimental data with MonteCarlo simulation for the IceCube 40-strings and AMANDA analysis. 
All error bands are purely statistical. After considering the systematic uncertainties described in the text, the declination 
distribution of the final neutrino sample is in agreement with the expectations of atmospheric neutrinos.}
\label{IC40_azimuth_declination}
\end{center}
\end{figure}

\begin{figure}[t]
\begin{center}
\includegraphics[width=0.4\textwidth]{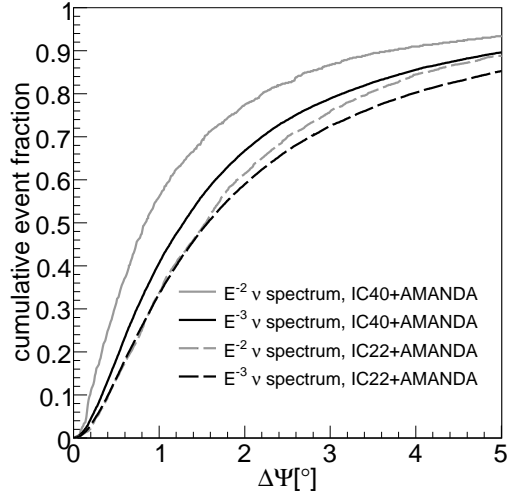}
\caption{Angular resolution of the two configurations IceCube-22 and IceCube-40 combined
with AMANDA. The angular resolution is determined for two different possible signal
spectra. The E$^{-2}$ neutrino spectrum includes on average events with higher
neutrino energies and better angular reconstruction. 
With the IC40+AMANDA configuration,
90\% of the neutrino events with an E$^{-2}$ neutrino spectrum are in the energy interval from 2.4~TeV to 750.0~TeV while the energy region
from 0.2~TeV to 20.5~TeV contains 90\% of the neutrino events if an E$^{-3}$ spectrum is considered.}
\label{ANG_RES}
\end{center}
\end{figure}

\begin{figure}[b]
\begin{center}
\includegraphics[width=3.5in]{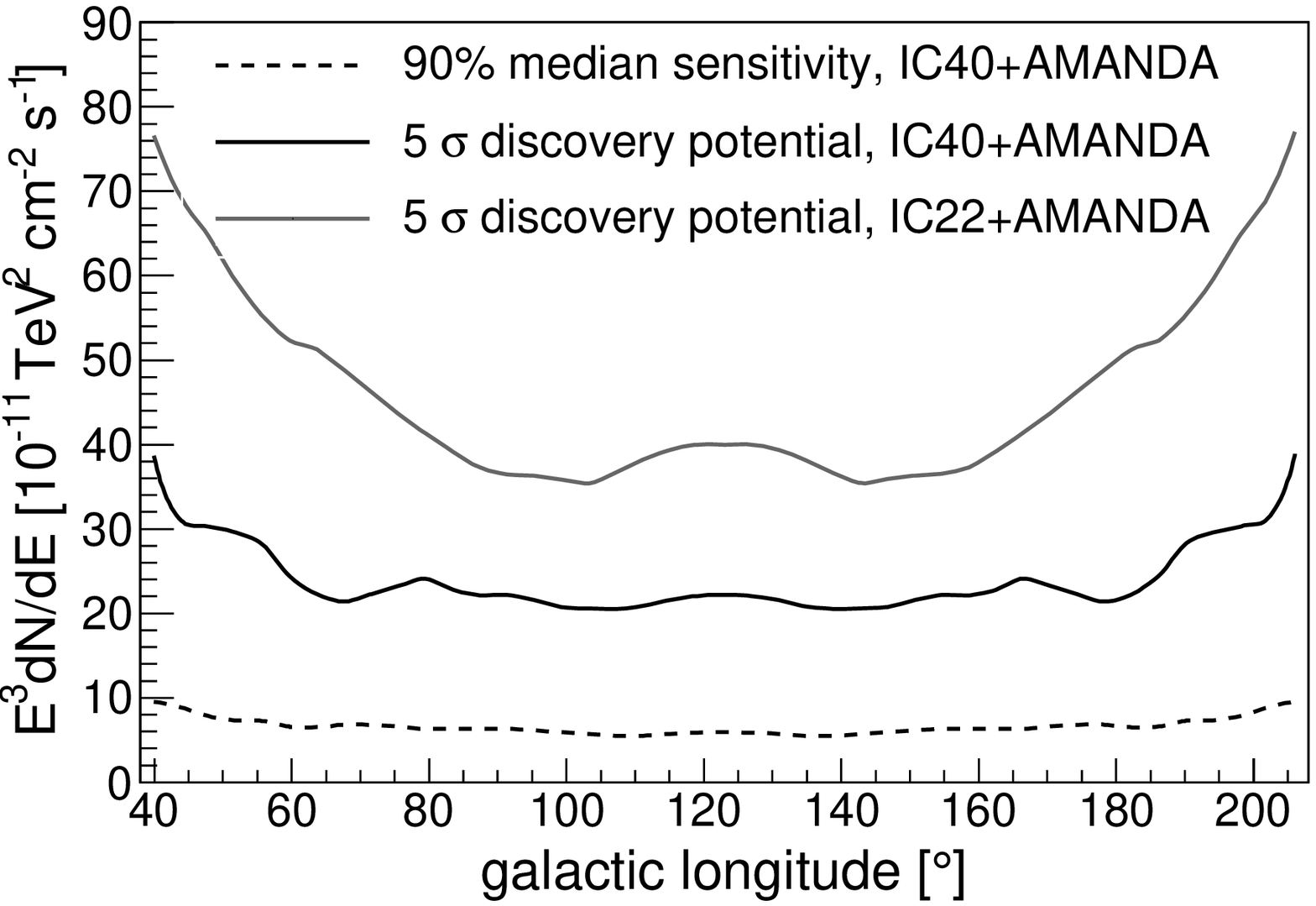}
\caption{Sensitivity and discovery potential 
of the IC40+AMANDA analysis for
E$^{-3}$ neutrino spectra. The sensitivity is calculated with the method
of Feldman and Cousins \citep{fel98} with 
a systematic uncertainty of $\pm 17\%$ on the neutrino flux 
included using the method defined in \citep{2003PhRvD..67a2002C} and modified in 
 \citep{2003PhRvD..67k8101H}.} 
\label{sensitivity}
\end{center}
\end{figure}

Figure~\ref{IC40_azimuth_declination} shows the distribution of the declinations of the events in the final neutrino sample.
The angular resolution obtained with this selection of events is shown in Figure~\ref{ANG_RES},
together with the angular resolution of the IC22+A analysis. The sensitivity and discovery potential for E$^{-3}$ neutrino spectra are reported in
Figure~\ref{sensitivity}.

\section{Results}
\label{Sec5}

As a result of the integrated use of the AMANDA 
detector within IceCube, we have obtained a significant improvement in 
the retention of neutrino induced events below a few TeV (see Figure~\ref{IC40_EFF_AREA_COMPARISON}).
This region is of importance for sources with soft or cutoff spectra. 
The tests described above resulted in no evidence for significant
deviation from the background-only hypothesis.
In the absence of detection of an astrophysical neutrino signal, upper limits on the muon neutrino flux from the considered regions of the Galaxy have been determined.
All the upper limits have been derived for soft neutrino spectra, with and without energy cutoff. 
The 90\% confidence level limits ($\Phi^{90\%}_{\nu_{\mu}}$) are calculated using the method of \citep{fel98}
i.e., $\mathrm{d\Phi_{\nu_{\mu}}/dE \leq \Phi^{90\%}_{\nu_{\mu}}~(E/TeV)^{-\alpha}~TeV^{-1}~cm^{-2}~s^{-1}}$. 
Systematic uncertainties have been included in the limit determination using the method defined in \citep{2003PhRvD..67a2002C} with the modification in \citep{2003PhRvD..67k8101H}. 

The upper limits are calculated for the total of the muon neutrino and
antineutrino flux reaching the Earth, assuming that no other neutrino
flavors contribute to the possible signal. For a source producing muon and electron neutrinos in the ratio of
about $\nu_{e}:\nu_{\mu}:\nu_{\tau} = 1:2:0$ typical of pion production from pp or p-gamma interaction,
neutrino oscillations with a large mixing angle $\theta_{23} \sim 45^{\circ}$ and long baseline result in approximate equipartition of flavors.
This analysis is to some extent sensitive to $\nu_{\tau}$ as well, 
mainly due to the decay of a $\tau$ into a $\mu$ with a branching ratio of $\approx 17\%$. Taking into account these effects and the details of energy losses, the contribution of $\nu_{\tau}$ 
is estimated to be an additional 10-16\% of the $\nu_{\mu}$ contribution for IceCube \citep{2011PhRvD..84h2001A} and AMANDA analyses \citep{abbasi09AMANDA}.\\
 
After completion of this analysis, a slight over-prediction of the muon neutrino flux
has been observed by comparing the results with an improved MonteCarlo simulation
not available before.
The intensity of the effect varies in declination and energy and it is estimated
to be less than 30\%. As a consequence, all upper limits reported in this work
as well as in \citep{2011ApJ...732...18A}, \citep{2011PhRvD..84h2001A} are slightly over-constraining.

\begin{figure}[t]
\begin{center}
\includegraphics[width=6.0in]{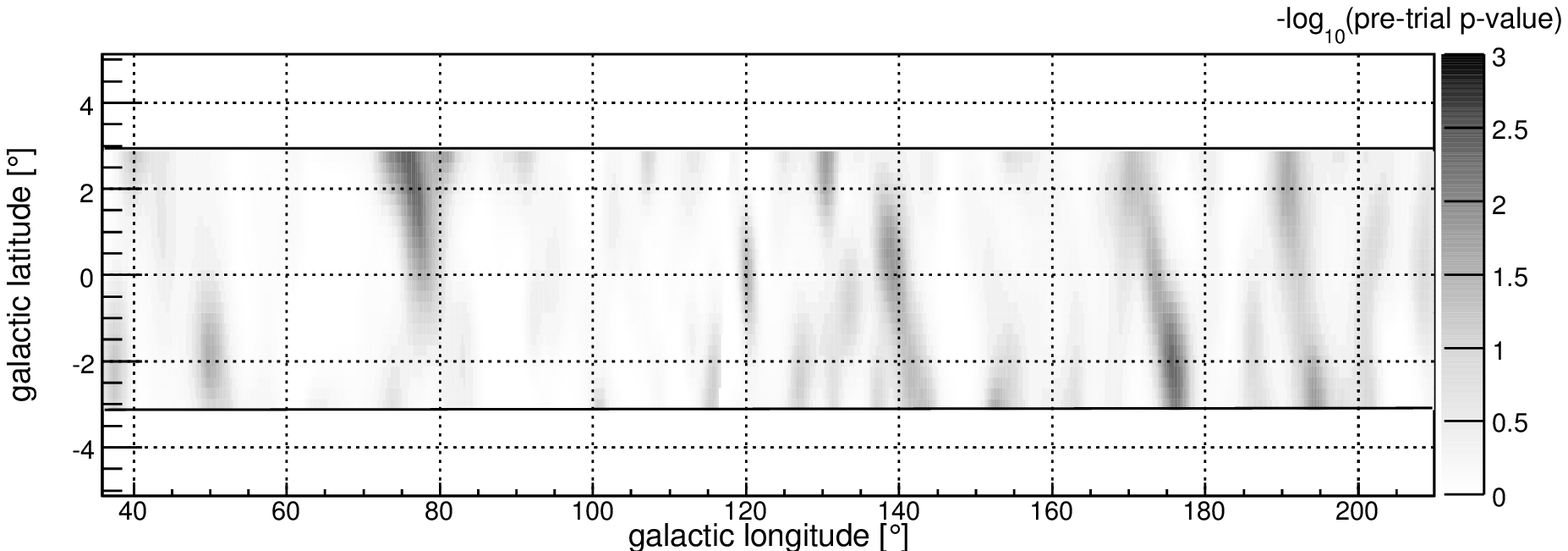}
\caption{Pre-trial significances (p-value) of the Galactic Plane scan using IceCube 22-strings and AMANDA data. }
\label{IC22AMA_ScanResult}
\end{center}
\end{figure}

\subsection{Galactic Plane scan and steady sources}
\begin{figure}[t]
\begin{center}
\includegraphics[width=6.0in]{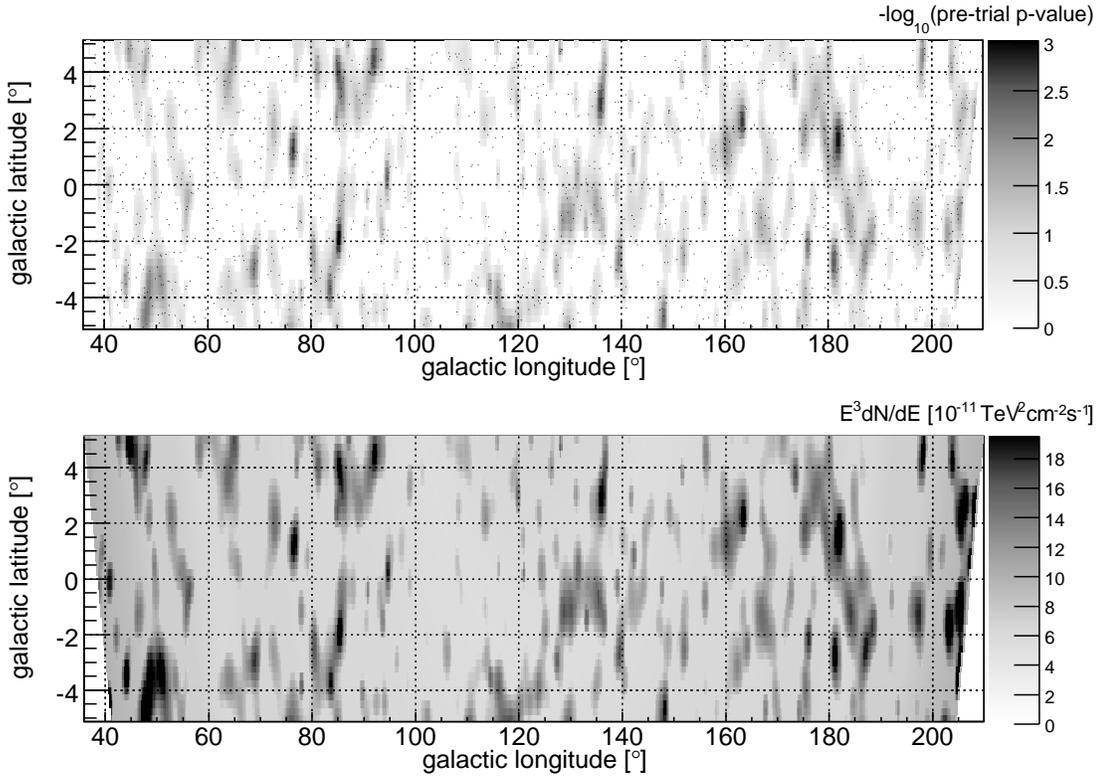}
\caption{Result of the scan of the Galactic Plane with data collected by IceCube 40-strings combined with AMANDA. 
Above: At each $0.25^{\circ} \times 0.25^{\circ}$ grid point, the result of the maximum likelihood analysis is shown in terms of its estimated significance ($-\log_{10}$ pre-trial p-value).
The reconstructed event locations are also indicated as black dots. 
The most significant excess of events is located at galactic coordinates $(85.5^\circ, -2,0^\circ)$ and has a pre-trial p-value of $0.09\%$ (-log(p-value)=3.03).
Accounting for the trials induced by the scanning of many points inside the Galactic Plane, the p-value of this test is $88\%$. Below: upper limits map for a
signal spectrum proportional to E$^{-3}$.} 
\label{IC40AMA_ScanResult}
\end{center}
\end{figure}

In Figures~\ref{IC22AMA_ScanResult} and \ref{IC40AMA_ScanResult}, we present the results of the scan of the Galactic Plane with IC22+A and with IC40+A. 
We note here that the considered region of the Galactic Plane covers only around 1/23 of the whole sky. This restriction in the tested area results in a lower trial factor compared to a $4\pi$-map.
Under the conservative assumption of a uniform angular resolution, the effective number of trials is expected to be a factor 23 lower than in an All-Sky scan like the one realized in \citep{2011ApJ...732...18A}. 
Assuming that the relation between the post-trial p-value $\rm{p_{post}}$ and the pre-trial p-value $\rm{p_{pre}}$ is $\rm{p_{post}=1-(1-p_{pre})^{N_{eff}}}$, the effective number of trials in the IC40+A
search is about $\rm{N_{eff}\approx2200}$. The reported post-trial p-values have been obtained by performing the analysis on randomized, and therefore signal-free, data samples.
 
In the analysis of IC22+A, the lowest background probability at point source angular scales is found at Galactic coordinates 
 $\ell = 75.9^\circ, b = 2.7^\circ$, with a pre-trial p-value of 0.37\%. Considering the intrinsic trials of the scan by analyzing randomized data samples,
an equal or higher significance in at least one of the scanned locations is found in 95\% of the cases.
The most significant point-like spot in the analysis of IC40+A, with a pre-trial p-value of 0.09\%, is found at Galactic coordinates $\ell = 85.5^\circ, b = -2.0^\circ$.
An equal or higher significance is found in $88\%$ of the randomized data samples. 

Upper limits on the neutrino emission have been calculated for the six
pre-selected candidate neutrino sources which have been studied for steady neutrino emission are summarized in 
Table~\ref{SourceListLimitsE3} for an E$^{-3}$ spectrum without cutoff. These limits have been obtained from the IC40+A analysis.

\begin{table}[t]
\begin{center}
\caption{\footnotesize E$^{-3}$ muon neutrino flux upper limits$^a$ from six $\gamma$-ray sources, based on IceCube 40-strings and AMANDA.} 
 \begin{tabular}{ l c c c c c c }
  \tableline\tableline
  Object & R.A. & Dec & $n_s$ & pre-trial p-value  & $\Phi^{90\%}_{\nu_{\mu}}$ \\
\tableline
    Crab Nebula& $83.63^\circ$ & $22.02^\circ$& 0 & - & 7.3 \\
    LSI +61 303& $40.13^\circ$ & $61.23^\circ$ & 1.6 & 0.25 & 8.3 \\
    W51 & $290.82^\circ$ & $14.15^\circ$ & 0.6 & - & 8.3   \\
    CasA& $350.85^\circ$ & $58.82^\circ$ & 0 & - & 5.9   \\
    SS433& $287.96^\circ$ & $4.98^\circ$ & 0 & - & 9.8  \\
    IC443& $94.18^\circ$ & $22.53^\circ$ & 0 & - & 7.3   \\
    \tableline
  \end{tabular}
  \tablenotetext{a}{The flux limits are given as $\Phi^{90\%}_{\nu_{\mu}}$ in units of
  $10^{-11}\mathrm{TeV^{-1} cm^{-2} s^{-1}}$ and represent the $90\%$ C.L. upper limit on the differential
  muon neutrino flux such that $d\Phi_{\nu_{\mu}}/dE \leq \Phi^{90\%}_{\nu_{\mu}} (\rm{E}/\mathrm{TeV})^{-3}$. p-values above 0.5 are given as ``-``.}
\label{SourceListLimitsE3}
\end{center}
\end{table}

In IC22+A, the highest excess in the candidate list was observed at the position of the Crab Nebula, with a pre-trial p-value of $13\%$ ($37\%$ post-trial). 
In IC40+A, the most significant clustering of events was observed for LSI +61 303. The 
estimated number (best-fit value) of signal events $n_{s}$ from this location is 1.6 and the observation corresponds to a pre-trial p-value of 25\%.
Accounting for the trials from testing six different positions, the post-trial p-value of this search is $42\%$, i.e.\ $42\%$ of randomized data samples show a similar or
stronger accumulation of events around one of the six objects. 

\subsection{The Cygnus region} 

\begin{figure*}[t]
\centering
\includegraphics[width=0.45\textwidth]{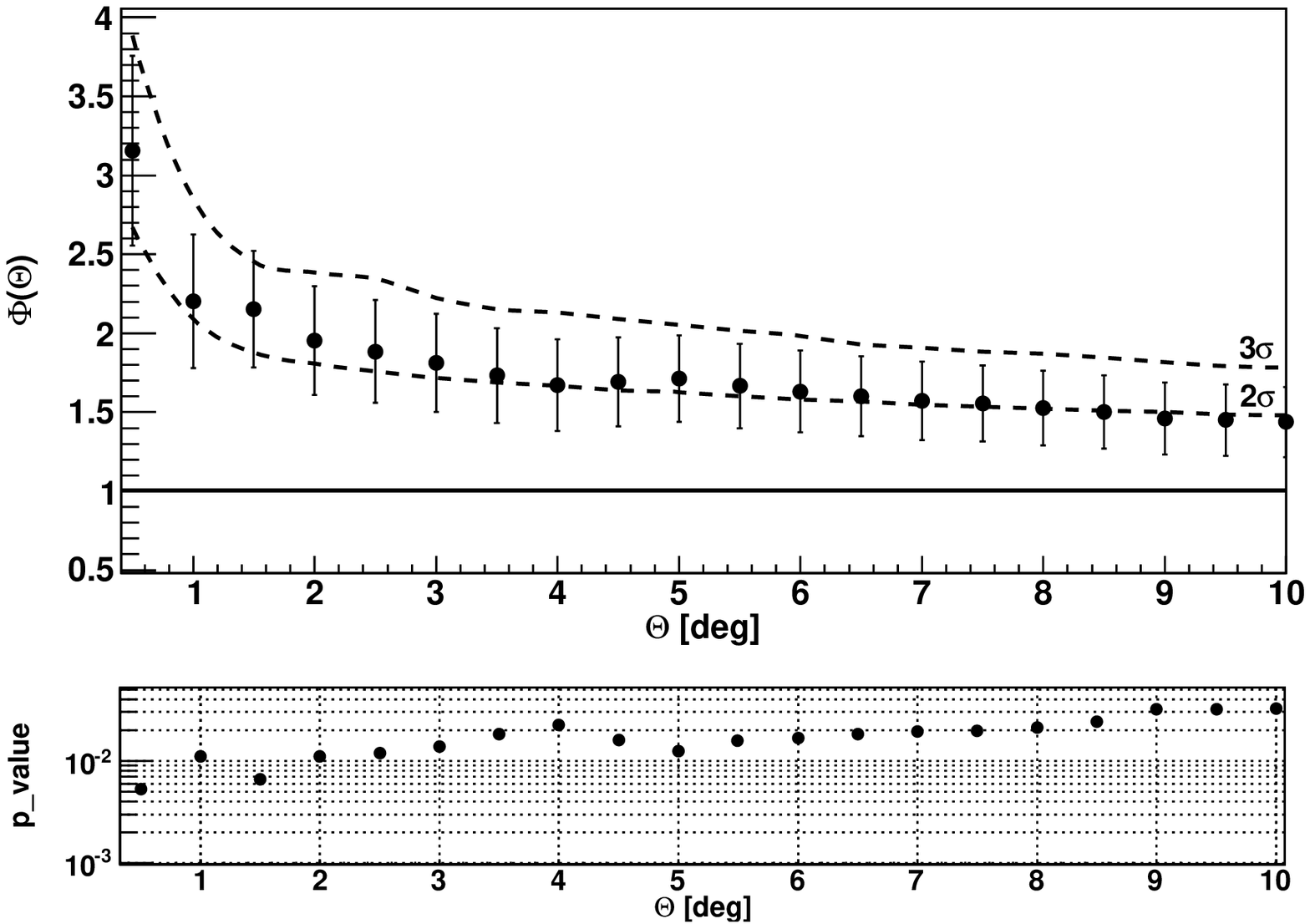}
\includegraphics[width=0.45\textwidth]{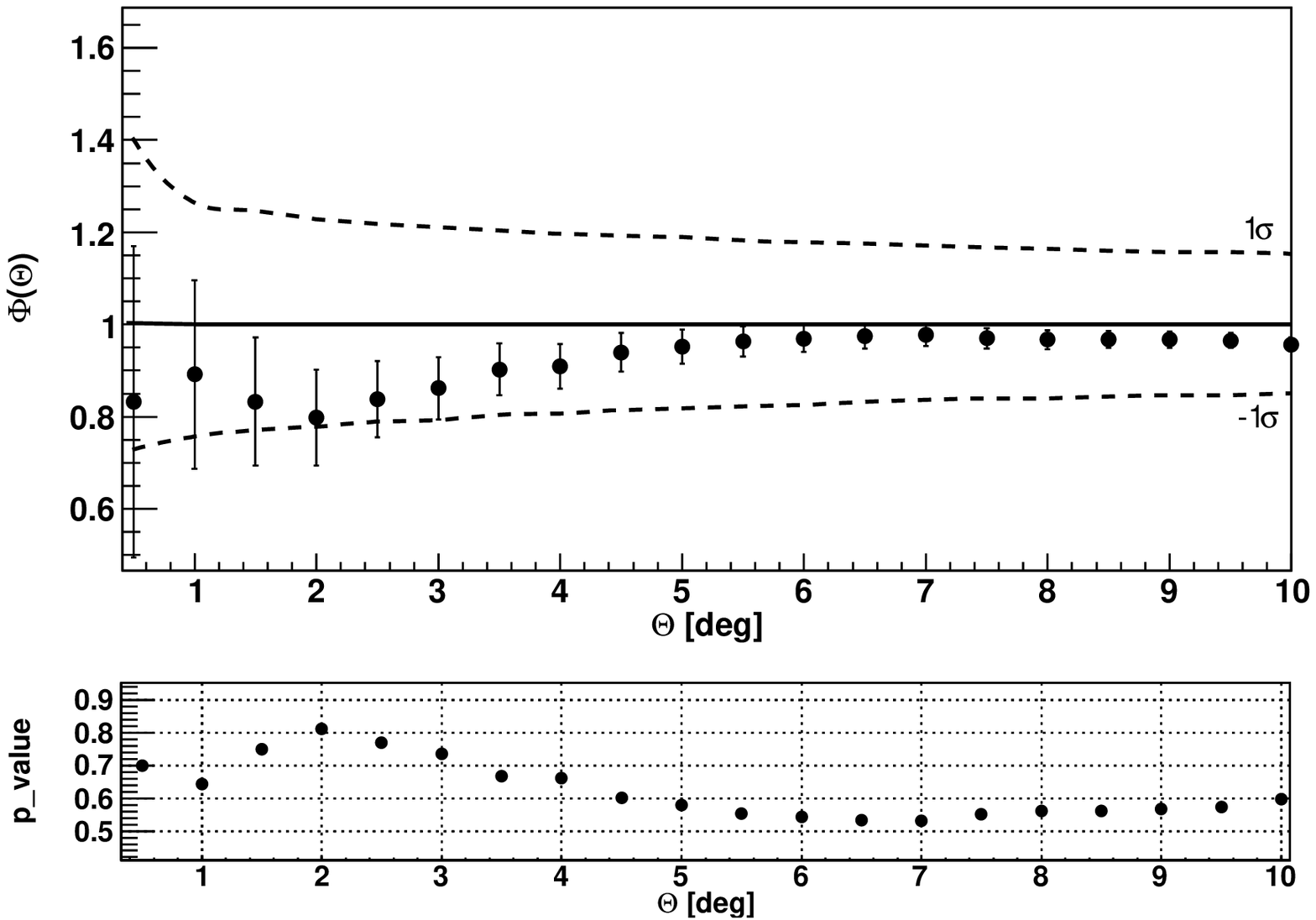}
\caption{
Measured clustering function of the neutrino events observed in the Cygnus region with IceCube 22-strings and AMANDA (left) and IceCube 40-strings and AMANDA (right).
Errors bars have been calculated by bootstrapping \citep{bootstrap}. In the left plot, the 2$\sigma$ and 3$\sigma$ levels are drawn with dashed lines and the p-value of the
clustering function at each of the angular scales tested is also shown. For IceCube 40-strings on the right, the 1$\sigma$ level is indicated by the dashed line.}
\label{Cluster}
\end{figure*}

\begin{figure*}[t]
\centering
\includegraphics[width=0.45\textwidth]{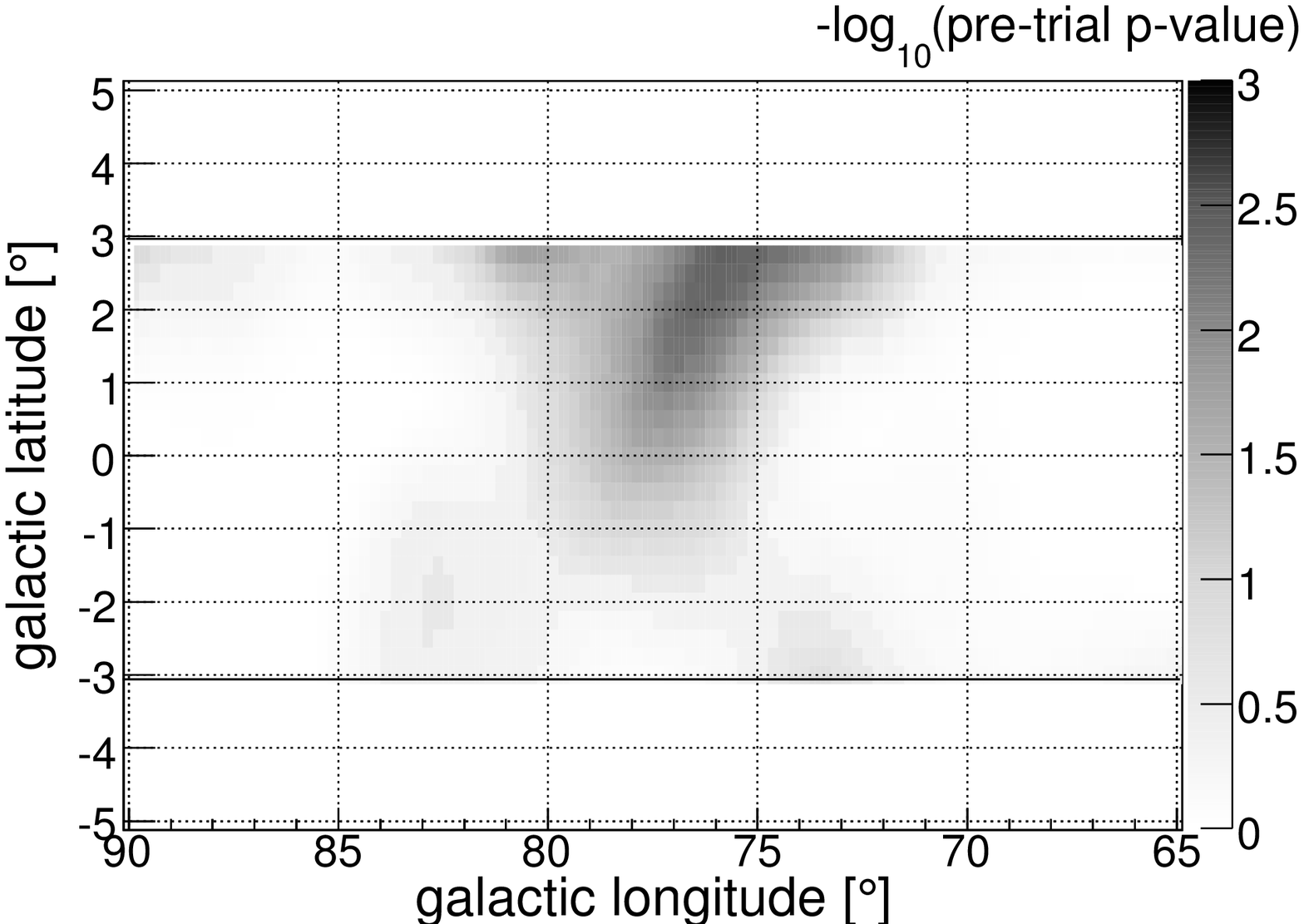} 
\includegraphics[width=0.45\textwidth]{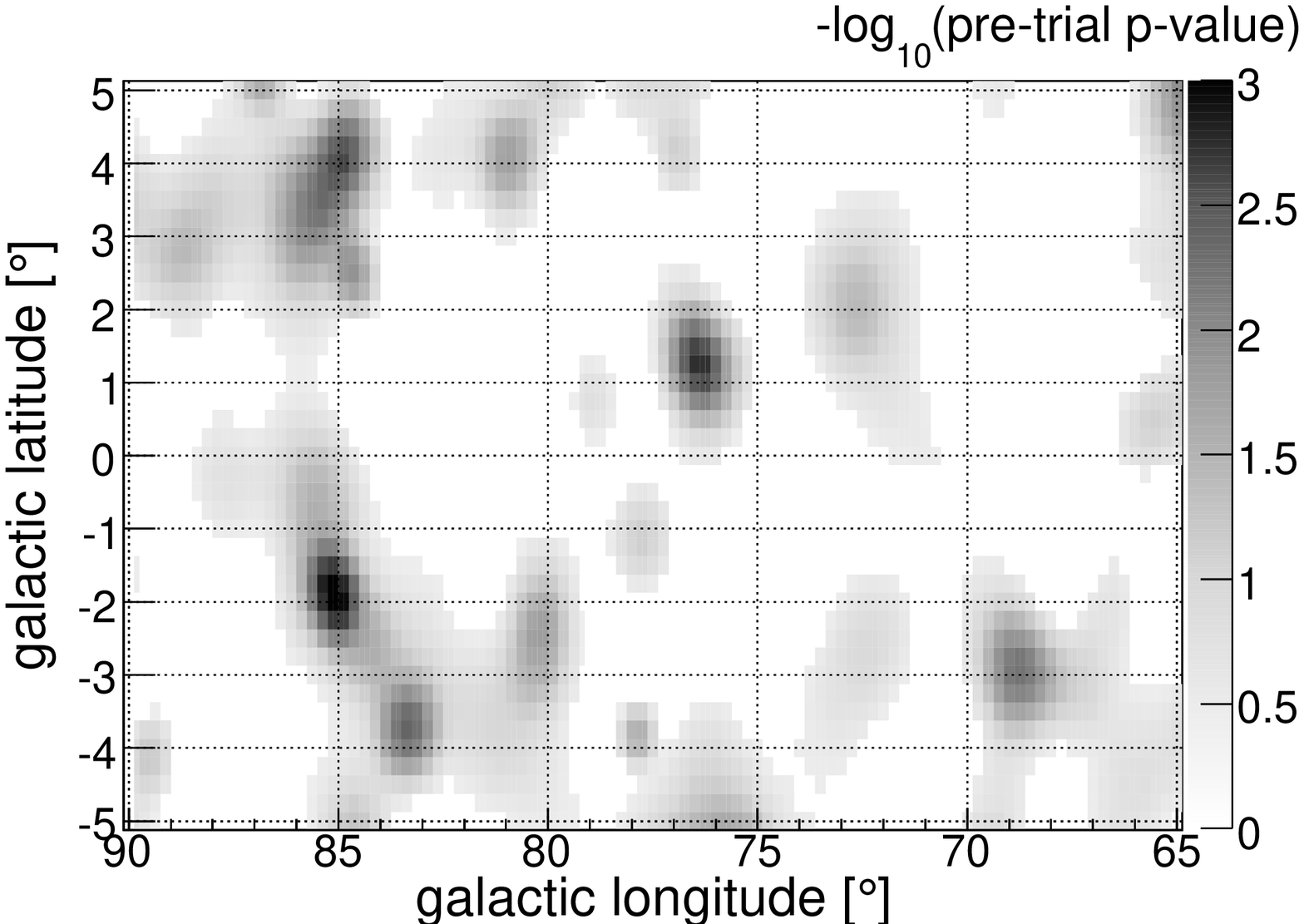}
\caption{A zoom in the Cygnus region obtained with the galactic plane scan with IceCube 22-strings and AMANDA (2007-2008) (left) and IceCube 40-strings and AMANDA (2008-2009) (right).
The gray shading indicates the negative logarithm of the p-value before trial correction.}
\label{CygnusIC22}
\end{figure*}

The clustering function $\Phi(\Theta)$ described in Sec.~\ref{method_cygnus_region} has been computed for events with galactic coordinates within $72^{\circ} < \ell < 83^{\circ}$ and $-3^{\circ} < b < 4^{\circ}$
for both IC22+A and IC40+A data. 
Figure~\ref{Cluster} shows the clustering function of the events in each sample; that is, the ratio of the number of event pairs separated by angular distance $\Theta$ or less,
with respect to the average number of such pairs for randomized events (i.e. the average case would therefore be located at $\Phi(\Theta)=1$). In the figure, we show the 2$\sigma$-3$\sigma$ (IC22+A sample)
and $\pm1\sigma$ (IC40+A) levels of the distribution of $\Phi(\Theta)$ under the hypothesis of a random distribution of events.
The observed values of $\Phi(\Theta)$ are represented together with measurement errors obtained by bootstrapping \citep{bootstrap}.
No significant concentration of events is seen at any of the angular scales tested. The result obtained on the 
IC22+A data sample shows a positive fluctuation 
at the level of 2.3~$\sigma$. This result contains already the correction of the trials obtained via scrambling and associated to the observation at different angular scales.
The region considered showed an excess with respect to the background expectation, which translates in excess values of $\Phi(\Theta)$ at all angular scales, but no significant structure is observed. 
The image of the Cygnus region obtained in IC22+A is shown in Figure~\ref{CygnusIC22} on the left.
However, the IC40+A analysis yields an under-fluctuation within the analyzed area, showing a rather dispersed distribution of events with respect to the average background case and its image
is in Figure~\ref{CygnusIC22} on the right.

The conclusions from both the Galactic Plane scans and the correlation analysis are therefore that the variations in the event density in the $11^{\circ} \times 7^{\circ}$ region
analyzed are consistent with background fluctuations. 

The under-fluctuation observed in IC40+A provides restrictive upper limits to point-like neutrino emission above 500~GeV from the Cygnus region. Upper limits have been computed 
from the measured value of the clustering function at $\Theta = 2^{\circ}$ using
a representative E$^{-2.6}$ power-law model 
under the assumption of a point-like neutrino signal located anywhere in the region.
That is, assuming that the spatial correlation of signal events in the region is given only by the PSF of the analysis.
The upper limits from the Cygnus region are at the level of $3\times 10^{-11}~\rm{TeV^{-1} cm^{-2} s^{-1}}$ for an E$^{-2.6}$ spectrum.

\subsection{Cygnus X-3} 
\begin{figure}[t]
\begin{center}
\includegraphics[width=0.7\textwidth]{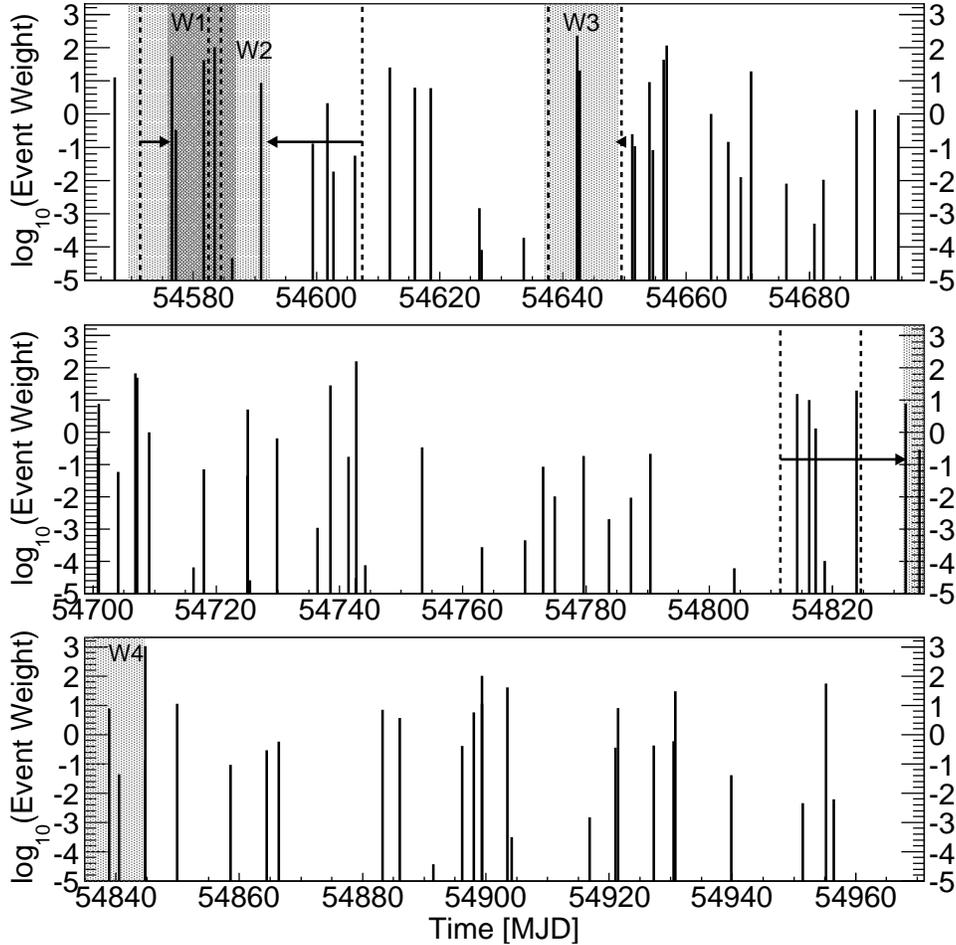}
\caption{Arrival times of neutrino events reconstructed within $10^\circ$ from Cyg X-3 and the position of the shifted search windows, each window being shifted individually.
The height of the line depicts the $\log_{10}$ of the spatial event weight (higher line means closer to the source). The dashed lines show the unshifted positions of the windows. The arrows indicate the shift.}
\label{fig:CygX-3-neutrino-times}
\end{center}
\end{figure}

In the flare search analysis of Cygnus X-3, 
no evidence for a signal is found in the neutrino sample for any of the sliding search windows. The smallest pre-trial 
p-value is $22\%$ (resulting from the search with window 1). After correction for the trials, we
arrive at a probability of $\approx\!57\%$ that this observation occurs in a background only sample
(final p-value). Upper limits on neutrino emission from Cyg X-3 during 20 days before
and after the time windows have been determined using the method proposed by Feldman \& Cousins
\citep{fel98} and are given in Table \ref{tab:UpperLimits}. Figure
\ref{fig:CygX-3-neutrino-times} shows the neutrino events close to Cyg X-3 as a function of time. 

\begin{table}[t!]
\begin{center}
\caption{\footnotesize Feldman-Cousins upper limits$^a$ on neutrino flux from Cyg~X-3, averaged over
 whole period of data-taking, assuming emission occurred only during the search windows (i.e. fluence divided by data-taking time).
Column 3 and 4 show the number of
events and expected number of background events within $5^\circ$
distance from assumed Cyg~X-3 position and within the shifted window
boundaries. $n_s$ is the best fit number of signal events from the
likelihood maximization.
 }
\begin{tabular}{cccccccc}
\tableline\tableline
Search & Shift & \multirow{2}{*}{Evts.} & Exp.     & \multirow{2}{*}{$n_s$}  & pre-trial & \multicolumn{2}{c}{$\Phi^{90\%}_{\nu_{\mu}}$} \\
Window & (days) &                        & bg. evts.  &                         & p-value   & E$^{-3}$ & E$^{-2}$ \\
     \tableline
       1 & $+4.46$    & $5$   & $2.6\pm1.3$   & $1.6$ & $0.22$ & $4.7$ & $0.73$ \\
       2 & $-15.05$   & $5$   & $4.6\pm1.9$   & $0.8$ & $0.42$ & $4.0$ & $0.60$ \\
       3 & $-0.58$    & $3$   & $3.0\pm1.5$   & $0.9$ & $0.39$ & $4.5$ & $0.59$ \\
       4 & $+20.00$   & $0$   & $3.0\pm1.5$   & $0.9$ & $0.24$ & $6.1$ & $0.77$ \\
       All & $+2.05$  & $12$  & $10.4\pm3.0$  & $1.0$ &  - & $5.0$ & $0.70$ \\
 \tableline
  \end{tabular}
  \tablenotetext{a}{The flux limits are given as $\Phi^{90\%}_{\nu_{\mu}}$ in units of
  $10^{-11}\un{TeV^{-1} cm^{-2} s^{-1}}$ which is the $90\%$ C.L. upper limit on the differential
 muon neutrino flux such that $d\Phi_{\nu_{\mu}}/dE \leq \Phi^{90\%}_{\nu_{\mu}} \cdot
 (\rm{E}/\mathrm{TeV})^{-\gamma}$, $\gamma=3$ or $2$ respectively. p-values above 0.5 are given as ``-``.}
   \label{tab:UpperLimits}
\end{center}
\end{table}

\section{Conclusions}
In this paper, we have presented dedicated searches for high-energy neutrino emission in 
the Galaxy. These analyses have been performed on data collected with two partial configurations 
of the IceCube neutrino telescope operating in conjunction with its predecessor AMANDA. In the two data taking periods 
considered here, IceCube operated in a 22-string and in a 40-string configuration, 
and AMANDA was an integrated part of IceCube in both seasons. 
Results from several searches have been presented. 
We have performed a scan of the Galactic Plane with the aim to discover point-like neutrino emission 
in the part of the Milky Way which is located in the northern hemisphere. 
Since no significant local clustering of events has been observed, upper limits for soft-spectra neutrino emission from 
the Galaxy have been reported.

A search that is sensitive for many possible morphologies of neutrino emission, 
including for example the presence of several weak or extended sources, has been performed for the Cygnus region of the Galactic plane, yielding restrictive upper limits. 
Both a strong TeV gamma-ray source (MGRO J2019+37) and a TeV diffuse component have been measured in the Cygnus region with the Milagro detector from the area defined
by Galactic latitude $ -3^{\circ} < b < 3^{\circ}$ and Galactic longitude $65^{\circ} < \ell < 85^{\circ} $ \citep{MgroCygnus}. The diffuse flux has been measured by Milagro over a region of
$\approx$0.02~sr and the total gamma-ray flux (diffuse and MGRO J2019+37) measured by Milagro accounts for $ \approx 10^{-11}$ TeV cm$^{-2}$ s$^{-1}$ at 12~TeV assuming
a differential source spectrum of E$^{-2.6}$. Under the hypothesis that the region is transparent to gamma-rays, the Milagro measurements can be used to estimate the maximal, associated neutrino flux
\citep{Kappes, 2006PhRvD..74c4018K}. Assuming that all the high-energy gamma-rays reported by Milagro come 
from decays of $\pi^{0}$ produced in proton-proton interactions, and using the same E$^{-2.6}$ spectrum
adopted in \citep{MgroCygnus}, the upper limits derived from the IC40+A analysis are only a factor of $\approx$2 \citep{Kappes, 2006PhRvD..74c4018K}
above this estimate of the maximal neutrino flux from inside the Cygnus region.
This implies that IceCube has the potential to detect neutrinos or to constrain 
the nature of the gamma-ray emission in one of the most active parts of the Galaxy in the next few years. 
Finally, a dedicated time-optimized search from the direction of the binary system Cygnus X-3 has been performed based on multi-wavelength observations. 
Upper limits for neutrino emission during specific episodes of enhanced radio and X-ray activity have been determined for this binary system. 

We have presented the first neutrino point source searches which use a more densely instrumented sub-array inside a large neutrino telescope. The capability to improve the performance in the energy range below $\sim$10~TeV in this way has been demonstrated. Using this capability, we have for the  first time optimized a search for neutrino point sources particularly for the more steeply falling energy spectra expected for Galactic neutrino sources.
 
AMANDA, the sub-array used in this work, has been decommissioned in 2009 and is now succeded by IceCube-DeepCore \citep{DCPaper}, an advanced low-energy extension of IceCube. The searches presented here have demonstrated that it is possible to improve the sensitivity to Galactic sources with early energy cut-offs and steeper spectra than E$^{-2}$ using a denser core array even at the expense of a larger atmospheric neutrino background. As a detector specifically built to enhance the sensitivity of IceCube at low energies, DeepCore is positioned in the deep center of the detector where the ice is clearest. This also allows to use the outer strings of IceCube as an atmosperic muon veto and thus to go beyond the approach taken in this work to improve the sensitivity in the energy range below $\sim$10~TeV.

\section*{Acknowledgments}
We acknowledge the support from the following agencies:
U.S. National Science Foundation-Office of Polar Programs,
U.S. National Science Foundation-Physics Division,
University of Wisconsin Alumni Research Foundation,
the Grid Laboratory Of Wisconsin (GLOW) grid infrastructure at the University of Wisconsin - Madison, the Open Science Grid (OSG) grid infrastructure;
U.S. Department of Energy, and National Energy Research Scientific Computing Center,
the Louisiana Optical Network Initiative (LONI) grid computing resources;
National Science and Engineering Research Council of Canada;
Swedish Research Council,
Swedish Polar Research Secretariat,
Swedish National Infrastructure for Computing (SNIC),
and Knut and Alice Wallenberg Foundation, Sweden;
German Ministry for Education and Research (BMBF),
Helmholtz Alliance for Astroparticle Physics (HAP),
Deutsche Forschungsgemeinschaft (DFG),
Research Department of Plasmas with Complex Interactions (Bochum), Germany;
Fund for Scientific Research (FNRS-FWO),
FWO Odysseus programme,
Flanders Institute to encourage scientific and technological research in industry (IWT),
Belgian Federal Science Policy Office (Belspo);
University of Oxford, United Kingdom;
Marsden Fund, New Zealand;
Australian Research Council;
Japan Society for Promotion of Science (JSPS);
the Swiss National Science Foundation (SNSF), Switzerland.

\end{document}